%% file: toi560.tex
\documentclass[fleqn,usenatbib]{mnras}

\usepackage{newtxtext,newtxmath}

\usepackage[T1]{fontenc}
\usepackage{fontawesome}

\DeclareRobustCommand{\VAN}[3]{#2}
\let\VANthebibliography\thebibliography
\def\thebibliography{\DeclareRobustCommand{\VAN}[3]{##3}\VANthebibliography}

\DeclareRobustCommand{\DA}[3]{#2}
\let\DAthebibliography\thebibliography
\def\thebibliography{\DeclareRobustCommand{\DA}[3]{##3}\DAthebibliography}

\usepackage{graphicx}	
\usepackage{amsmath}	

\newcommand{\pyaneti}{\href{https://github.com/oscaribv/pyaneti}{\texttt{pyaneti}\,\faGithub}}
\newcommand{\logr}{$\log R'_{\rm HK}$}
\newcommand{\sshk}{$S_{\rm HK}$}
\newcommand{\lbe}{$\lambda_{\rm e}$}
\newcommand{\lbp}{$\lambda_{\rm p}$}
\newcommand{\pgp}{$P_{\rm GP}$}
\newcommand{\gcm}{${\rm g\,cm^{-3}}$}
\newcommand{\ms}{${\rm m\,s^{-1}}$}
\newcommand{\kms}{${\rm km\,s^{-1}}$}
\newcommand{\gs}{${\rm g\,s^{-1}}$}
\newcommand{\citlalicue}{\texttt{citlalicue}}

\newcommand{\vsini}{$v \sin i$}
\newcommand{\logg}{$\log g$}

\newcommand{\msun}{$M_{\odot}$}
\newcommand{\rsun}{$R_{\odot}$}

\newcommand{\feh}{\ensuremath{[\mbox{Fe}/\mbox{H}]}}
\newcommand{\teff}{\ensuremath{T_{\mathrm{eff}}}}
\newcommand{\vmic}{\ensuremath{v_{\mathrm{mic}}}}
\newcommand{\vmac}{\ensuremath{v_{\mathrm{mac}}}}

\newcommand{\tess}{\emph{TESS}}
\newcommand{\cheops}{\emph{CHEOPS}}
\newcommand{\toi}{TOI-560}
\newcommand{\hd}{HD\,73583}
\newcommand{\hdb}{HD\,73583\,b}
\newcommand{\hdc}{HD\,73583\,c}
\newcommand{\hdbc}{HD\,73583\,b and c}

\input{TOI-560-full_params}

\title[The young HD 73583 (TOI-560) planetary system]{The young HD 73583 (TOI-560) planetary system: Two 10-M$_\oplus$ mini-Neptunes transiting a 500-Myr-old, bright, and active K dwarf}

\author[O. Barrag\'an et al.]{
O.~Barragán$^{1}$\thanks{\texttt{oscar.barrragan@physics.ox.ac.uk}, \href{https://twitter.com/oscaribv}{\faTwitter\texttt{@oscaribv}}} , 
D.~J.~Armstrong$^{2,3}$, 
D.~Gandolfi$^{4}$, 
I.~Carleo$^{5}$, 
A.~A.~Vidotto$^{6}$, 
C.~Villarreal~D'Angelo$^{7}$, \and  
A.~Oklop\v{c}i\'c$^{8}$, 
H.~Isaacson$^{9}$, 
D.~Oddo$^{10}$
% ...
\\  \\
Affiliations appear at the end of the paper
}

\date{Accepted XXX. Received YYY; in original form ZZZ}

\pubyear{2015}

\begin{document}
\label{firstpage}
\pagerange{\pageref{firstpage}--\pageref{lastpage}}
\maketitle

\begin{abstract}
We present the discovery and characterisation of two transiting planets observed by \textit{TESS} in the light curves of the young and bright (V=9.67) star HD73583 (TOI-560). We perform an intensive spectroscopic and photometric space- and ground-based follow-up in order to confirm and characterise the system. We found that HD73583 is a young ( {$\sim 500$~Myr}) active star with a rotational period of \jPGP[]\,d, and a mass and radius of \smass\ and \sradius, respectively. \hdb\ ($P_b=$\Pb) has a mass and radius of \mpb\ and \rpb, respectively, that gives a density of \denpb. \hdc\ ($P_c$=\Pc) has a mass and radius of \mpc\ and \rpc, respectively, this translates to a density of \denpc. 
Both planets are consistent with worlds made of a solid core surrounded by a volatile envelope. 
Because of their youth and host star brightness, they both are excellent candidates to perform transmission spectroscopy studies. 
 { We expect  on-going atmospheric mass-loss for both planets caused by stellar irradiation. We estimate that the detection of evaporating signatures on H and He would be challenging, but doable with present and future instruments.}
\end{abstract}

\begin{keywords}
Planets and satellites: individual: HD\,73583 (TOI-560) -- Stars: activity -- Techniques: photometric -- Techniques: radial velocities.
\end{keywords}



\section{Introduction}

Two of the most noticeable characteristics of the transiting exoplanet population are the so-called ``hot Neptunian desert'' \citep{Mazeh2016,Lundkvist2016} and the ``radius valley'' \citep{Fulton2017,VanEylen2018}.
Both correspond to regions with a lack of planets within certain ranges of planetary radii and stellar irradiance. 
Theoretical evolution models suggest that these gaps are mainly caused by physical mechanisms that occurs during the  {first Myr of evolution ($\lesssim 1$\,Gyr)}, such as photo-evaporation \citep[e.g.,][]{Adams2006,Kubyshkina2018b,Lopez2014,Mordasini2020,Raymond2009,Owen2013} and core-powered mass-loss \citep[e.g.,][]{Ginzburg2016,Gupta2019,Gupta2021}.

Young exoplanets ($<1$\,Gyr) offer us snapshots of early planetary evolution that can be used to test the role of diverse physical mechanisms sculpting exoplanet populations. 
The few well-characterised young exoplanets have given us some insights on the role of photo-evaporation in early exoplanet evolution.
One of them is K2-100\,b {($\sim 750$ Myr)}, a young and highly irradiated exoplanet that lies on the border of the hot Neptunian desert \citep{Mann2017}.
Recent studies suggest that the planet is currently evaporating and its radius will be significantly smaller in a few Gyr, which will eventually cause it to leave the hot Neptunian desert \citep{Barragan2019}.
Another example is the AU\,Mic\,b {($\sim 22$ Myr)} planet \citep{Cale2021,Klein2021,Plavchan2020,Szabo2021}, whose density is consistent with a planet with a thick volatile envelope that may be evaporating \citep[e.g.,][]{Carolan2020}. 

The Transiting Exoplanet Survey Satellite \citep[\tess;][]{Ricker2015} 
has discovered a plethora of candidates/exoplanets transiting bright stars \citep[e.g.,][]{Bouma2020,Hobson2021,Kossakowski2021,Martioli2021,Mann2021,Newton2019,Newton2021,Plavchan2020,Rizutto2020,Zhou2021}. 
These transiting exoplanets are excellent targets to perform follow-up observations that allow us to further characterise these young systems, e.g., using the radial velocity (RV) method to measure the planetary masses. 
However, detecting the planetary signatures in RV time-series of young stars is challenging due to their inherent activity.
Active regions on stellar surfaces induce an apparent RV shift that can mimic or hide planetary signals \citep[e.g.,][]{Faria2020,Huelamo2008,Queloz2001,Rajpaul2016}.
Therefore, in order to detect the planetary signals in stellar RVs, state-of-the-art spectrographs are not enough and we need to use techniques tailored to disentangle planetary and stellar signals in our RV data \citep[see e.g.,][]{Barragan2018b,Grunblatt2015,Hatzes2010,Hatzes2011,Haywood2014,Rajpaul2015}.
This is especially important for exoplanets with expected RV signals of the order of a few \ms, which is similar to (or smaller than) the stellar signals of some  young active stars \citep[e.g.,][]{Barragan2019,Kossakowski2021,Lillo2020}.

In this paper we present the discovery and mass measurement of two mini-Neptunes transiting \hd\  (\toi, TIC~101011575), a young star observed by \tess\ in Sectors 8 and 34. \hd\ is a relatively bright ($V=9.67$) and high proper motion star located in the southern hemisphere. Table~\ref{tab:parstellar} shows the main identifiers for \hd.
The characterisation of this system is part of the \emph{KESPRINT} \citep[e.g.,][]{Carleo2020,Esposito2019,Gandolfi2018,Gandolfi2019,Georgieva2021} and \emph{NCORES} \citep[e.g.][]{Armstrong2020,Nielsen2020,Osborn2021} consortia, that have discovered and characterised several \tess\ exoplanets. 
This manuscript is organised as follows: In Section~\ref{sec:tess} we describe the \tess\ observations. Section~\ref{sec:followup} is devoted to the description of our intensive photometric and spectroscopic follow-up of the star. Sections~\ref{sec:stellar} and \ref{sec:dataanalysis} describe our stellar and planetary data analyses, respectively. We close in Section~\ref{sec:discusion} with our discussion and conclusions.

\begin{table}
\caption{Main identifiers, coordinates, proper motion, parallax, and optical and infrared magnitudes of \hd.  \label{tab:parstellar} 
}
\begin{center}
\begin{tabular}{lcc} 
\hline
\hline
\noalign{\smallskip}
Parameter & Value &  Source \\
\noalign{\smallskip}
\hline
\noalign{\smallskip}
\multicolumn{3}{l}{\emph{Main identifiers}} \\
\noalign{\smallskip}
TIC & 101011575   & TIC$^{(a)}$ \\
Gaia DR2  & 5746824674801810816 & TIC$^{(a)}$, Gaia$^{(b)}$  \\
TYC & 5441-00431-1 & TIC$^{(a)}$ \\
2MASS & J08384526-1315240 & TIC$^{(a)}$ \\
Spectral type & K4V & \citet{Gray2006} \\
\noalign{\smallskip}
\hline
\noalign{\smallskip}
\multicolumn{3}{l}{\emph{Equatorial coordinates, proper motion, and parallax}} \\
\noalign{\smallskip}
$\alpha$(J2000.0) &  08 38 45.26042 & TIC$^{(a)}$, Gaia$^{(b)}$ \\
$\delta$(J2000.0) & -13 15 24.0910  & TIC$^{(a)}$, Gaia$^{(b)}$ \\
$\mu_\alpha$\,(mas\,${\rm yr^{{-1}}}$) & $-63.8583 \pm 0.050515$ & TIC$^{(a)}$, Gaia$^{(b)}$  \\
$\mu_\delta$\,(mas\,${\rm yr^{{-1}}}$) & $38.3741 \pm 0.040586$ &  TIC$^{(a)}$, Gaia$^{(b)}$ \\
$\pi$\,(mas) & $ 	31.6501 \pm 0.0319$ & \ Gaia$^{(b)}$  \\
Distance\,(pc) & $31.60 \pm 0.032$ & This work \\
\noalign{\smallskip}
\hline
\noalign{\smallskip}
\multicolumn{3}{l}{\emph{Magnitudes}} \\
\tess &	$8.5925 \pm 0.006$ &TIC$^{(a)}$ \\ 
Gaia & $9.27033 \pm 0.00048 $ &	TIC$^{(a)}$, Gaia$^{(b)}$  \\ 	  		
B &	$10.742 \pm 0.07$ &	TIC$^{(a)}$ \\  		
V &	$9.67 \pm 0.03$ &	TIC$^{(a)}$  \\ 	 		
J & $7.649 \pm 0.026$ &
	\citet{Cutri2003} 	 \\ 	  		
H &	$7.092 \pm 0.051$ &
	\citet{Cutri2003} 	 \\ 	  		
Ks &	$6.948 \pm 0.024$ &
	\citet{Cutri2003} 	 \\ 	  		
W1 & 	$6.85 \pm 0.037$ &
	\citet{wise} 	 \\ 	  		
W2 &	$6.963 \pm 0.021$ &
	\citet{wise} 	 \\ 	  		
W3 	& $6.921 \pm 0.017$ &
	\citet{wise}  	 \\ 	  		
W4 	& $6.723 \pm 0.084$ &
	\citet{wise}  	 \\ 	 
\noalign{\smallskip}
\hline
\multicolumn{3}{l}{\footnotesize$^a$ TESS Input Catalog \citep[TIC;][]{Stassun2018,Stassun2019}.}\\
\multicolumn{3}{l}{\footnotesize$^b$ \citet{Gaia2018}.}\\
\end{tabular}
\end{center}
\end{table}


\section{\tess\ photometry}
\label{sec:tess}

\begin{figure*}
    \centering
    \includegraphics[width=\textwidth]{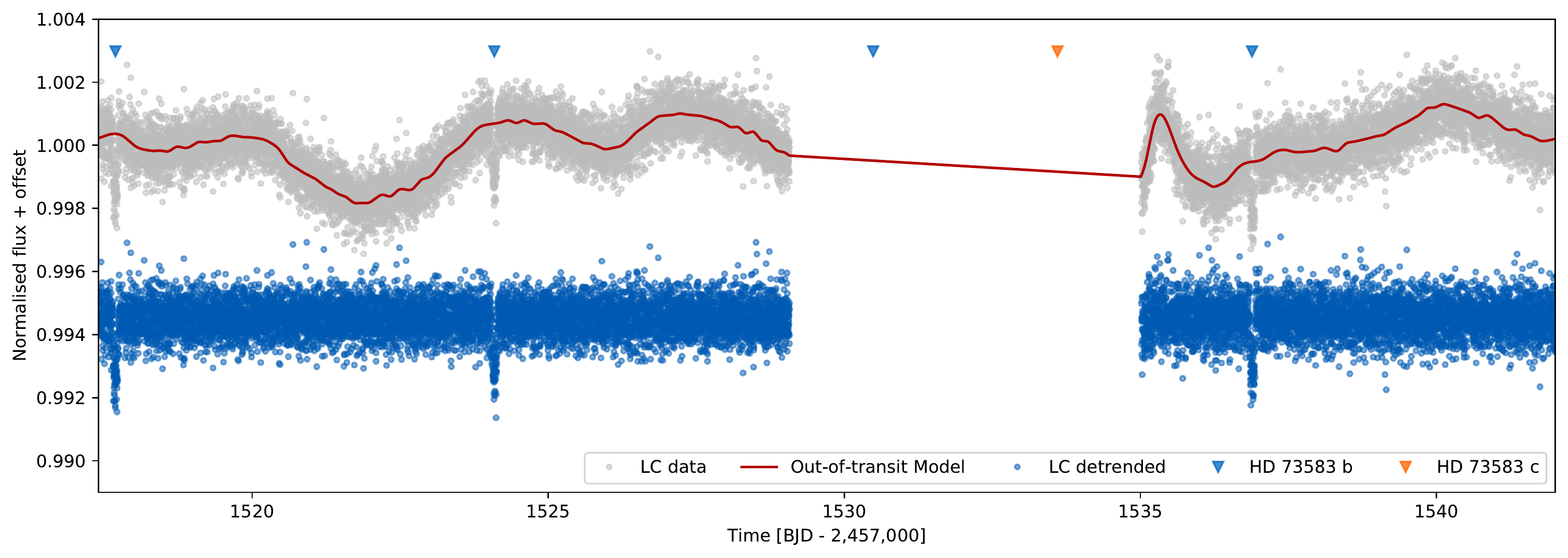}\\
    \includegraphics[width=\textwidth]{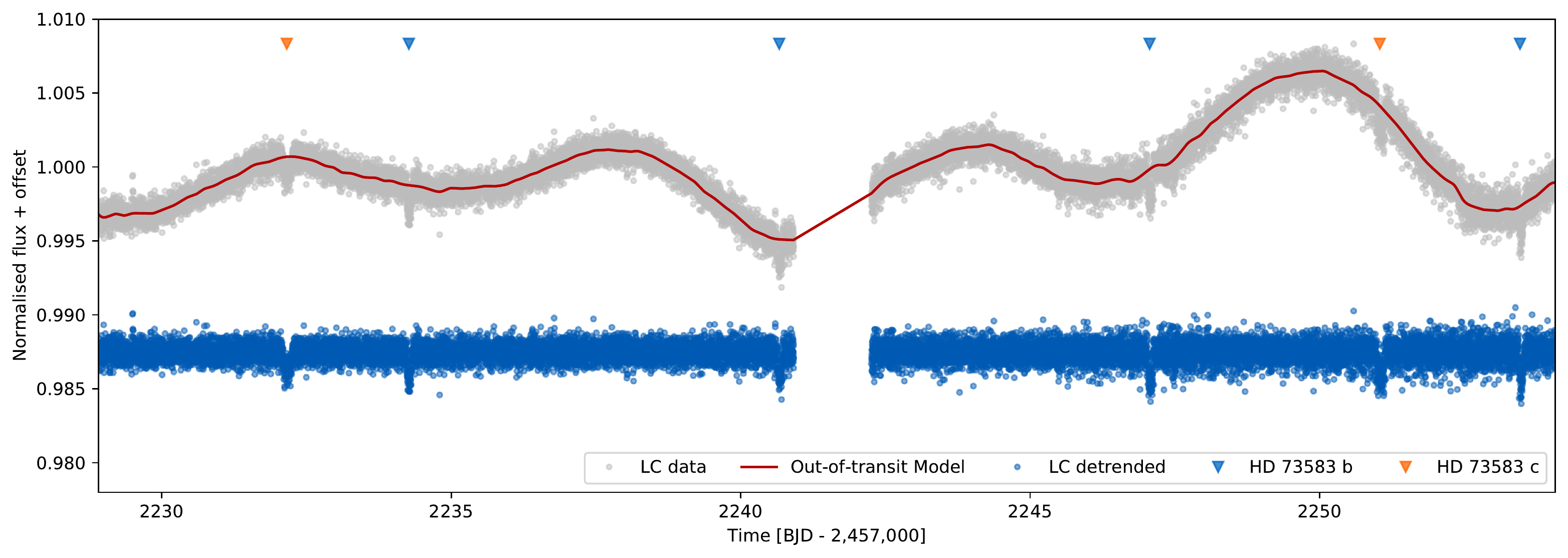}
    \caption{\tess's Sector 8 (upper panel) and Sector 34 (lower panel) light curves for \hd.
    \tess\ data are shown with grey points with the out-of-transit variability model over-plotted in red. 
    The resulting flattened light curves are shown with blue points.
    Transit positions are marked with blue and orange triangles for \hdb\ and \hdc, respectively.}
    \label{fig:lcs}
\end{figure*}

\tess\ observed \hd\ (\toi, TIC~101011575) in Sector 8 from 2019 February 02 to 2019 February 28 on camera 2 with a cadence of 2-min. 
The \tess\ Science Processing Operations Center \citep[SPOC;][]{jenkins2016} transit search \citep{Jenkins2002,Jenkins2010,Jenkins2020} discovered a transiting signal with a period of 6.4~d in \hd’s light curve.
This was announced in the \tess\ SPOC Data Validation Report \citep[DVR;][]{Twicken2018,Li2019}\footnote{DVR for \hdb\ can be found in this \href{https://mast.stsci.edu/api/v0.1/Download/file/?uri=mast:TESS/product/tess2019033200935-s0008-s0008-0000000101011575-00182_dvm.pdf}{link}.}, and designated by the TESS Science Office as TESS Object of Interest \citep[TOI;][]{Guerrero2021} TOI-560.01 (hereafter \hdb).
We identified \hd\ as a good candidate and we started an intensive follow-up to further characterise the nature of this system (See Sect.~\ref{sec:followup}). 
Figure~\ref{fig:lcs} shows the sector 8 normalised Presearch Data Conditioning Simple Aperture Photometry \citep[PDCSAP;][]{Smith2012,Stumpe2012,Stumpe2014} light curve for \hd\ as downloaded from the Mikulski Archive for Space Telescopes (MAST).
We note that Sector 8 has a relatively large data gap of more than 5 days. This was caused by an interruption in communications between the instrument and spacecraft that resulted in no collection of data during this time\footnote{Sector 8 release notes can be found in this \href{https://ntrs.nasa.gov/api/citations/20190002566/downloads/20190002566.pdf}{link}.}. 

Two years later, \tess\ re-observed \hd\ as part of its extended mission in Sector 34 between 2021 January 13 and 2021 February 09 in camera 2 with a 2-min cadence.
We downloaded the Sector 34 \hd's light curve from the MAST archive. We found that the expected transit signals associated with \hdb\ were consistent with the transits observed in Sector 8.
We also detected by eye two extra transits in Sector 34 that do not have a counterpart in Sector 8 at times 2232.17 and 2251.06 BTJD, {where BTJD $=$ BJD - 2\,457\,000 is the Barycentric \tess\ Julian date}. 
These two new transits have similar depths ($\sim 1130$\,ppm), suggesting that they are caused by the same transiting object with a period of 18.9\,d. 
At this point, we had enough spectroscopic data to test a planetary origin of the signal. We performed a preliminary analysis and detected a Doppler signal in our RVs consistent with the 18.9-day period (See Sect.~\ref{sec:dataanalysis} for the full details on the RV analysis), suggesting that these transits have a planetary origin.
Hereafter we refer to this signal as \hdc. The reason why there is no \hdc's transit signal in Sector 8 is because the expected transit time coincides with the relatively long data gap (see Fig.~\ref{fig:lcs}).
\hdc\ was also announced as a Community TOI (CTOI) and TOI in the EXOFOP website as \toi.02. Both TOIs were detected with the correct ephemerides in the SPOC transit search of the combined data for sectors 8 and 34 \citep{Guerrero2021}.

We note that the Sector 34 PDCSAP light curve shows significantly systematic variations, especially during the second half of the observations. According to \tess' Sector 34 release notes, orbit 76 suffered from  significant spacecraft motion\footnote{Sector 34 release notes can be found in this \href{https://tasoc.dk/docs/release_notes/tess_sector_34_drn50_v02.pdf}{link}.}.
This suggests that the apparent PDCSAP light curve corruption is likely caused by an over-fitting of the Cotrending Base Vectors (CBVs) when trying to correct the significant spacecraft motion.
For this reason we decided to perform our own light curve correction using the \texttt{lightkurve} software \citep{lightkurve}.
Briefly, we use the \texttt{CBVCorrector} class to perform a `Single-scale' and `Spike' CBV correction. We first set a regularisation term \texttt{alpha}\,$=1\times10^{-4}$. This produces a corrected light curve that is visibly similar to the PDCSAP one with an Over-fitting metric of 0.5 that is smaller than the recommended threshold of 0.8\footnote{For more details about CBV correction of \tess\ data see \url{https://docs.lightkurve.org/tutorials/2-creating-light-curves/2-3-how-to-use-cbvcorrector.html}.}.
We therefore perform a scan over different values of the regularisation term that provides an optimal Over and Under-fitting metric.
We found that the best regularisation term is \texttt{alpha}\,$=9.4\times 10^3$. This implies a small correction of the original light curve by the CBVs, and the corrected data is practically identical to the SAP light curve.
We therefore perform a `Spike' only CBV correction (to only correct for short impulsive spike systematics) with a a regularisation term \texttt{alpha}\,$=1\times10^{-4}$. This generates a corrected light curve with an Over-fitting metric of 0.83 and Under-fitting metric of 0.91. These values are above the recommended values, and we therefore use this as our corrected light curve for Sector 34. 
To finish the light curve processing we performed a crowding correction to account for extra flux that may be present in the SAP mask. We use the values given in the target pixel file to account and correct for the light curve contamination of $\sim$4\%\footnote{See~\url{https://heasarc.gsfc.nasa.gov/docs/tess/UnderstandingCrowding.html} for more details on \tess\ crowding correction.}. 
Figure~\ref{fig:lcs} shows our processed Sector 34 light curve for \hd.

Astrophysical and instrumental false positives are very common in \tess\ data, in part due to the large pixel scale of 21 arcseconds. We, therefore, performed standard diagnostic tests to help rule out false positive scenarios using the open source Lightcurve Analysis Tool for Transiting Exoplanets \citep[\texttt{LATTE};][]{latte}. 
The tests to check for instrumental false positives include ensuring that the transit events do not coincide with the periodic momentum dumps and assessing the $x$ and $y$ centroid position around the time of the events. 
Similarly, tests for astrophysical false positives include assessing the background flux; examining the light curves of nearby stars observed by \tess; examining light curves extracted for each pixel around the target; and comparing the average in-transit with the average out of-transit flux. 
These tests increased our confidence that none of the transit-signals are the result of systematic effects, such as a temperature change in the satellite, or that they are astrophysical false positives such as background eclipsing binaries or a solar system object passing through the field of view.

\hd's \tess\ light curves show out-of-transit variability likely caused by activity regions on the stellar surface and/or instrumental systematics. 
For further transit analysis in this manuscript, we chose to remove the low-frequency trends in order to work with flattened light curves.
We detrended the \tess\ light curves using the public code \href{https://github.com/oscaribv/citlalicue}{\texttt{citlalicue} \faGithub} \citep{pyaneti2}. Briefly, \citlalicue\ uses Gaussian Processes (GPs)
as implemented in \texttt{george} \citep[][]{george} to model the out-of-transit variability in the light curves.
We fed \citlalicue\ with the normalised light curves and we input the ephemeris of the two transiting signals. Since we are interested in removing the low frequency signals, we bin the data to 3 hours bins and mask out all the transits from the light curve when fitting the GP using a Quasi-Periodic kernel {\citep[as described in][]{george}}. 
We use an iterative maximum Likelihood optimisation together with a $5$-sigma clipping algorithm to find the optimal model describing the out-of-transit light curve variations. 
We then divide the whole light curves by the inferred model to obtain a flattened light curve containing only transit signals. 
We note that we detrended each \tess\ sector independently. 
Figure~\ref{fig:lcs} shows the detrended light curves for both \tess\ sectors.
In Sect.~\ref{sec:dataanalysis} we present the modelling of the flattened \tess\ transits.

\section{Follow-up observations}
\label{sec:followup}

\subsection{High Resolution Speckle Imaging}
\label{sec:speckle}

Spatially close stellar companions can create a false-positive transit signal if, for example, the fainter star is an eclipsing binary (EB). However, even more troublesome is ``third-light'' flux contamination from a close companion (bound or line of sight) which can lead to underestimated derived planetary radii if not accounted for in the transit model \citep[e.g.,][]{Ciardi2015} and even cause total non-detection of small planets residing within the same exoplanetary system \citep{Lester2021}. 
Thus, to search for close-in bound companions to exoplanet host stars that are unresolved in \tess\ or other ground-based follow-up observations, we obtained high-resolution imaging speckle observations of \hd.

\hd\ was observed twice, on 2020 March 16 and 2019 May 22 UT, using the Zorro speckle instrument on the Gemini South 8-m telescope\footnote {\url{https://www.gemini.edu/sciops/instruments/alopeke-zorro/}}.
The March 2020 observations will be discussed herein as the May 2019 observations had poorer seeing and worse sky conditions, however giving similar results to those obtained in March 2020. Zorro provides simultaneous speckle imaging in two bands (562 nm and 832 nm) with output data products including a reconstructed image with robust contrast limits on companion detection \citep[e.g.,][]{Howell2016}. 

\begin{figure}
    \centering
    \includegraphics[width=0.48\textwidth]{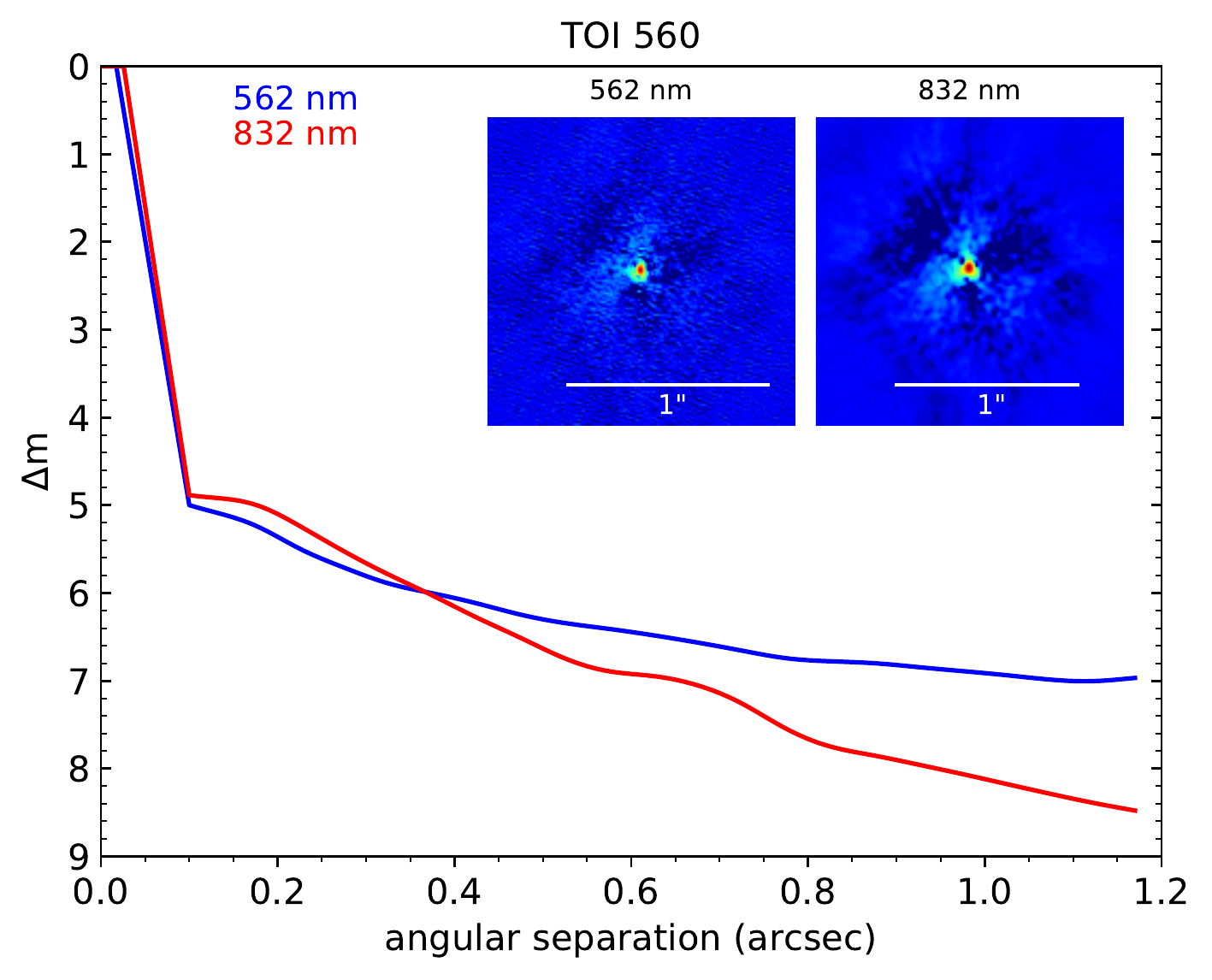}
    \caption{Contrast curves showing the $5\,\sigma$ detection sensitivity obtained using Zorro on Gemini with the filters centred on 562nm (blue line) and 832nm (red line). No bright companions are detected within 1.2 arcsec to \hd.}
    \label{fig:speckle}
\end{figure}

Three sets of $1000 \times 0.06$ s exposures were collected for TOI 560 and subjected to Fourier analysis in our standard reduction pipeline \citep[see][]{Howell2011}. Figure~\ref{fig:speckle} shows our final 5-sigma contrast curves and the reconstructed speckle images.
We find that \hd\ is single to within the contrast limits achieved by the observations, with no companion brighter than 5-8 magnitudes below that of the target star found from the diffraction limit (20 mas) out to 1.2~arcsec.
At the distance of \hd\ these angular limits correspond to spatial limits of 0.6 to 37 AU. 

\subsection{\cheops\ observations}
\label{sec:cheops}

We observed \hd\ with the CHaracterising ExOPlanet Satellite \citep[\cheops;][]{Benz2021} as part of our Guest Observer program (OBS ID 1345790) between 2021-01-26 UTC 01:40 and 2021-01-26 UTC 06:39, so as to capture a full transit of \hdb\ with its precise photometry.
\cheops\ is an ESA small-class mission with ultra-high-precision photometry dedicated to follow-up of stars with known planets. 
It conducts observations from a Sun-synchronous, low-Earth orbit, which in many cases leads to interruptions in on-target observations due to Earth occultations and crossings of the South Atlantic Anomaly (SAA) region. As such, each \cheops\ observation is associated with some observing efficiency, which is the fraction of time on target not interrupted by Earth occultation and SAA. 
We obtained one visit on the target at a relatively high efficiency of 72.7\% for a duration of 4.99 hours. 
\cheops\ observations were then passed through the CHEOPS Data Reduction Pipeline \citep[DRP;][]{Hoyer2020}, which conducts calibration, correction, and photometry, as shown in the Data Reduction Report for each observation. 
We chose to use the ``OPTIMAL'' lightcurve as provided by the DRP, which calculates optimal aperture size by maximising the SNR based on field-of-view simulations to account for potential contaminant field stars. In this case, the  ``OPTIMAL'' aperture size was 26.0 pix, whereas the ``DEFAULT'' aperture size was 25 pix.

The light curve was then detrended using the \href{https://github.com/pmaxted/pycheops}{\texttt{pycheops} \faGithub} Python package (Maxted et al., submitted). 
We detrended the \cheops\ light curve to remove spacecraft motion, background noise and Moon glint. In an effort to avoid overfitting, we tested whether the addition of a new detrending parameter was supported by the data by calculating the Bayes Factor of the model with/without the parameter one-by-one, and eliminating those parameters whose Bayes Factors $\geq 1$.
We use this detrended light curve for our analyses described in Sect.~\ref{sec:dataanalysis}.

\subsection{Ground-based transit observations}
\label{sec:fut}

We conducted ground-based photometric follow-up observations as part of the {\em TESS} Follow-up Observing Program \citep[TFOP;][]{collins:2018}. 
We used the {\tt TESS Transit Finder}, which is a customised version of the {\tt Tapir} software package \citep{Jensen:2013}, to schedule our transit observations. A summary of the observations is provided in Table \ref{table:SG1-phot-obs}.

We observed a full transit event of \hdb\ simultaneously using four Next Generation Transit Survey \citep[NGTS;][]{wheatley18ngts} 0.2\,m telescopes located at ESO's Paranal Observatory, Chile. Each NGTS telescope has an 8 square degree field-of-view and a plate scale of 5\,arcsec\, pixel$^{-1}$.  The observations were taken using a custom NGTS filter (520 - 890nm) with 10\,s exposure times and at airmass < 1.95.  Each telescope independently observed the transit event in the multi-telescope operational mode as described in \citet{bryant20multicam} and \citet{Smith2020}.  The NGTS data were reduced using a custom aperture photometry pipeline \citep{bryant20multicam}, which uses the SEP library for both source extraction and photometry \citep{bertin96sextractor, Barbary2016}.

We observed a full transit of \hdb\ from the Perth Exoplanet Survey Telescope (PEST) near Perth, Australia. The 0.3 m telescope is equipped with a $1530\times1020$ SBIG ST-8XME camera with an image scale of 1$\farcs$2 pixel$^{-1}$ resulting in a $31\arcmin\times21\arcmin$ field of view. A custom pipeline based on {\tt C-Munipack}\footnote{\url{http://c-munipack.sourceforge.net}} was used to calibrate the images and extract the differential photometry.

We observed three full transits of \hdb\ and one ingress and two simultaneous egresses of \hdc\ from the Las Cumbres Observatory Global Telescope \citep[LCOGT;][]{Brown:2013} 1.0\,m network.  The $4096\times4096$ LCOGT SINISTRO cameras have an image scale of $0\farcs389$ per pixel, resulting in a $26\arcmin\times26\arcmin$ field of view. The images were calibrated by the standard LCOGT {\tt BANZAI} pipeline \citep{McCully:2018}, and photometric data were extracted with {\tt AstroImageJ} \citep{Collins:2017}.

\begin{table}
\centering
\caption{Summary of Ground-based Photometric Follow-up Observations}
\label{table:SG1-phot-obs}
    \begin{tabular}{llllll}
    \hline\hline
    Telescope & Location & Date & Filter  & Coverage  \\
           &           & [UTC]&               &           \\
    \hline
{\hdb}\\
\hline
NGTS 0.2\,m         & Chile              & 2019-12-06   &  NGTS        & full \\
PEST 0.3m       &  Australia   & 2020-01-14   &  $\mathrm{R_c}$         &  full  \\
LCO-SAAO 1m     & South Africa       & 2020-02-02   &  $z$-short   &  full  \\
LCO-SSO 1m      & Australia          & 2020-03-31   &  B           & full  \\
LCO-SAAO 1m     & South Africa       & 2020-12-05   &  $z$-short   &  full  \\
\hline
{\hdc}\\
\hline
LCO-CTIO 1m & Chile      & 2021-04-03   &  $z$-short   &  ingress \\
LCO-McDonald 1m & U.S.A.      & 2021-04-22   &  $z$-short   &  egress \\
LCO-McDonald 1m & U.S.A.      & 2021-04-22   &  $z$-short   &  egress \\
\hline
\end{tabular}
\end{table}


\subsection{Radial velocity follow-up}

\subsubsection{HARPS}

We acquired 90 high-resolution (R\,$\approx$\,115\,000) spectra with the High Accuracy Radial Velocity Planet Searcher \citep[HARPS;][]{Mayor2003} spectrograph mounted at the 3.6\,m ESO telescope at La Silla Observatory. 
HARPS observes in the visible spectrum within the wavelength range of 380 to 690 nm.
The typical exposure time per observation was 1500 s, this produced spectra with a typical S/N of 70-80 at 550 nm.
The observations were carried out between April 2019 and March 2020, as part of the two large observing programs 1102.C-0923 (PI: Gandolfi) and 1102.C-0249 (PI: Armstrong), and the ESO programs 60.A-9700 and 60.A-9709. We reduced the data using the dedicated HARPS data reduction software (DRS) and extracted the radial velocity (RV) measurements by cross-correlating the Echelle spectra with a K5 numerical mask \citep{Baranne1996,Pepe2002, Lovis2007}. We also used the \texttt{DRS} to extract the Ca\,{\sc ii} H\,\&\,K lines activity indicator (\sshk), and three profile diagnostics of the cross-correlation function (CCF), namely, the contrast, the full width at half maximum (FWHM), and the bisector inverse slope (BIS).
{Our HARPS RV measurements have a typical error bar of 1.2\,\ms\ and a RMS of 9.1\,\ms.}
Table~\ref{tab:harps} lists the HARPS RV and activity indicators measurements.

We acknowledge that there are eight archival HARPS observations of \hd\ taken in 2004 and 2005 (program: 072.C-0488, PI: Mayor). We note that the stellar activity may have changed significantly in the last 15 years, and that those observations have a sub-optimal sampling. Therefore, we do not include them in our time-series analysis.

\subsubsection{PFS}

\citet{Teske2020} performed spectroscopic follow-up observations of \hd\ as part of the Magellan-TESS Survey, which uses the Planet Finder Spectrograph \citep[PFS;][]{Crane2010} on the 6.5\,m Magellan II telescope at Las Campanas Observatory in Chile. PFS covers wavelengths from 391 to 734 nm.
Twenty-six high resolution (R\,$\approx$\,130\,000) spectra were acquired between 08 April 2019 and 24 May 2019 UT. 
The typical integration time for each observation was 1200 s.
{These RVs measurements have a typical error bar of 0.8\,\ms\ and a RMS of 6.3\,\ms.}
We decided to include the PFS measurements in our spectroscopic time-series analysis given that they overlap with our HARPS observations. The Table containing the Doppler and S-index measurements are available in electronic format in \citet{Teske2020}.

\subsubsection{HIRES}

We collected fourteen iodine-in observations and two template observations of HD 73583 between 2019 Oct 20 and 2020 Jan 01. 
Each spectrum was taken with the B5 decker with width 0.86" and height of 3.5", resulting in resolution of approximately 60\,000. The iodine-out observations that serve as in the RV forward model consist of two back to back observations of SNR $\sim200$ each. The median observation time of the iodine in observations is 560 seconds resulting in SNR $\sim300$ at 550 nm, the middle of the iodine cell absorption region. 
The setup and data reduction follow the standard procedure laid out in \citet{Howard2010}. RVs have a median internal error of 1.0 \ms\ and pre-fit RMS of 12.2 \ms.  
The RV errors from HIRES for this young star are higher than they would be for a solar-aged star. 
Table~\ref{tab:hires} shows the HIRES RV and \sshk\ measurements.

\subsubsection{CORALIE}

\hd\ was observed with the CORALIE high resolution echelle spectrograph on the 1.2 m Euler telescope at La Silla Observatory~\citep{CORALIE}. The star was part of a blind RV survey for planets around K-dwarfs within 65~pc. In total 17 spectra were obtained between 2016-09-24 and 2017-12-04 UT. One observation was discarded from further analysis due to abnormal instrument drift during the exposure.
RVs were extracted via cross-correlation with a binary G2 mask \citep{Baranne1996}, using the standard CORALIE data-reduction pipeline. We also derived cross-correlation (CCF) line-diagnostics such as bisector-span and FWHM to check for possible false-positive scenarios \citep{Queloz2001}. {We obtained a typical SNR of 50 that, {together with intrinsic signals in the data}, gives a RV precision of 5-7~\ms and a RMS of the measurements of 12~\ms}. The Ca II index was computed using the usual prescription. Table~\ref{tab:coralie} shows the CORALIE spectroscopic observations.

\section{Stellar data analysis}
\label{sec:stellar}

\subsection{Spectroscopic parameters}

We used the HARPS observations to produce a high signal-to-noise (S/N=700 at 550\,nm) spectrum of \hd. 
Briefly, we arbitrarily chose one of the RV observations as the zero velocity and through a cross-correlation procedure shift and add all the other observations to this one. 
Because of the inherent lack of accuracy in {\it all} available methods, we use three independent methods in order to derive the basic three spectroscopic parameters: effective temperature (\teff), surface gravity (\logg), and metallicity (\feh). 

The first method, using \texttt{SpecMatch-Emp} \citep{Yee2017}, compares a standardised version of our spectrum to a library of more than 400 spectra of stars with well determined parameters. Through interpolation and a minimising process the code provides a set of stellar fundamental parameters such as \teff, metallicity (\feh), \logg.
{\texttt{SpecMatch-Emp} also provides estimates of the stellar mass and radius by comparing with stellar masses and radii for stars in the sample during the minimisation process.}
Table~\ref{tab:stellarparams} shows the main spectroscopic parameters for \hd\ obtained with \texttt{SpecMatch-Emp}.

Our second method is utilising the code \texttt{SME} \citep[Spectroscopy Made Easy;][]{Piskunov2017}, a well proven tool for determining stellar parameters using synthetic spectra.
By providing some basic information about the observed spectrum, like estimates of the fundamental stellar parameters like \teff, \logg, \feh, \vsini, \vmac\ or \vmic\ together with atomic or molecular line data from the  VALD3 database \citep{Piskunov1995,Kupka1999}, one can fit the observed spectrum. 
Calling a dynamically linked external library of models, \texttt{SME} performs a synthesis of the stellar atmospheric spectrum. Functions in the library solve for molecular and ionisation equilibrium, continuous and line opacities, calculating spectra while solving for the parameters left free. Using an iterative scheme, varying one or two fundamental parameters at a time, and with the inherent chi-square minimising technique, \texttt{SME} eventually arrives at the most appropriate parameters for the model best fitting the observed spectrum.
Table~\ref{tab:stellarparams} shows the main spectroscopic parameters obtained with \texttt{SME}. 

Our third method uses \texttt{ARES+MOOG}, following the same methodology described in \citet[][]{Santos-13, Sousa-14, Sousa-21}. 
We used the combined spectrum to derive the equivalent widths (EW) of iron lines using the \texttt{ARES} code\footnote{The last version of ARES code (ARES v2) can be downloaded at \url{http://www.astro.up.pt/$\sim$sousasag/ares}} \citep{Sousa-07, Sousa-15}. 
We used a minimisation process to find ionisation and excitation equilibrium and converge to the best set of spectroscopic parameters. This process makes use of a grid of \citet{Kurucz-93} model atmospheres  and the radiative transfer code \texttt{MOOG} \citep{Sneden-73}. 
Following the same methodology as described in \citet[][]{Sousa-21}, we used the GAIA eDR3 gaia paralax and estimated the trigonometric surface gravity to be 4.60 $\pm$ 0.06 dex. Table~\ref{tab:stellarparams} shows a summary with our \texttt{ARES+MOOG} results. 

\subsection{Stellar mass and radius}
\label{sec:stellarmassradius}

Our three methods to retrieve stellar spectroscopic parameters provide results that agree well within the uncertainties (Table~\ref{tab:stellarparams}). We decide to adopt the \texttt{SpecMatch-Emp} parameters for further analyses given that this method provides the most conservative error bars.
We then use the \texttt{SpecMatch-Emp} spectroscopic stellar parameters together with \texttt{PARAM\,1.3}\footnote{\url{http://stev.oapd.inaf.it/cgi-bin/param_1.3}.} \citep{daSilva2006} with the \texttt{PARSEC} isochrones \citep{Bressan2012} to derive \hd's mass and radius.
We use the \teff\ and \feh\ from our \texttt{SpecMatch-Emp} analysis together with the visual magnitude and parallax given in Table~\ref{tab:parstellar} as input for \texttt{PARAM\,1.3}.
{Given the expected youth of the star, we set stellar age priors between 0.1 and 1 Gyr (See Sect.~\ref{sec:stellarage}).}
Table~\ref{tab:stellarparams} shows \hd's parameters obtained with \texttt{PARAM\,1.3}. We note that these parameters are in full agreement within 1-sigma with the mass and radius estimated using \texttt{SpecMatch-Emp}.

\begin{table*}
\centering
\caption{\hd's parameters. The adopted parameters for the rest of the manuscript are marked with boldface.}
\label{tab:stellarparams}
    \begin{tabular}{lcccccccc}
    \hline\hline
    Parameter & \texttt{SpecMatch-Emp} & \texttt{SME} & \texttt{ARES+MOOG} & \texttt{PARAM\,1.3} &
    \texttt{ARIADNE} & \texttt{stardate} & time-\logr & WASP \\
    \hline
    \teff\ (K) & $\mathbf{4511 \pm 110}$ & $4532 \pm 80 $ & $4555 \pm 99$ & $\cdots$  &  $\cdots$  & $\cdots$  &  $\cdots$ &  $\cdots$ \\
    \logg\ (cgs) & $\mathbf{4.62 \pm 0.12}$ & $4.42 \pm 0.08$ & $4.60 \pm  0.06$ & $4.63 \pm 0.02$  & $\cdots$  & $\cdots$   &  $\cdots$  &  $\cdots$\\
    \feh\ (dex) & $\mathbf{0.00 \pm 0.09}$ & $0.00 \pm 0.05$  & $-0.132    \pm 0.053$ & $\cdots$  & $\cdots$ &  $-0.02 \pm 0.06$  &  $\cdots$ &  $\cdots$ \\
    \vsini\ & $\cdots$ & $3.5 \pm 0.5$ & $\cdots$  & $\cdots$ & $\cdots$ & $\cdots$  &  $\cdots$ &  $\cdots$\\
    Mass (\msun) & $0.72 \pm 0.08 $ & $\cdots$ & $\cdots$ &  $\mathbf{0.73 \pm 0.02}$ & $0.727^{+0.022}_{-0.030}$ & $0.74 \pm 0.02$  &  $\cdots$ &  $\cdots$\\
    Radius (\rsun) & $0.70 \pm 0.10$ & $\cdots$ & $\cdots$ & $\mathbf{0.65 \pm 0.02}$   & $0.693^{+0.019}_{-0.030}$  & $\cdots$  &  $\cdots$ &  $\cdots$\\
    Density (\gcm) & $2.9_{-1.0}^{+1.8}$ & $\cdots$ & $\cdots$ & $\mathbf{3.75 \pm 0.38}$  & $3.08^{+0.46}_{-0.39}$ & $\cdots$  &  $\cdots$ &  $\cdots$\\
    Age (Gyr) & $\cdots$ & $\cdots$ & $\cdots$ & $0.44 \pm 0.33$ & $\cdots$ & $0.750 \pm 0.020$   &  $\mathbf{0.48 \pm 0.19}$ &  $\cdots$\\
    Rotation period (d) & $\cdots$ & $\cdots$ & $\cdots$ & $\cdots$ & $\cdots$ & $\cdots$ &  $\cdots$ & $12.2 \pm 0.2 $ \\
\hline
\end{tabular}
\end{table*}

\begin{figure}
    \centering
    \includegraphics[width=0.48\textwidth]{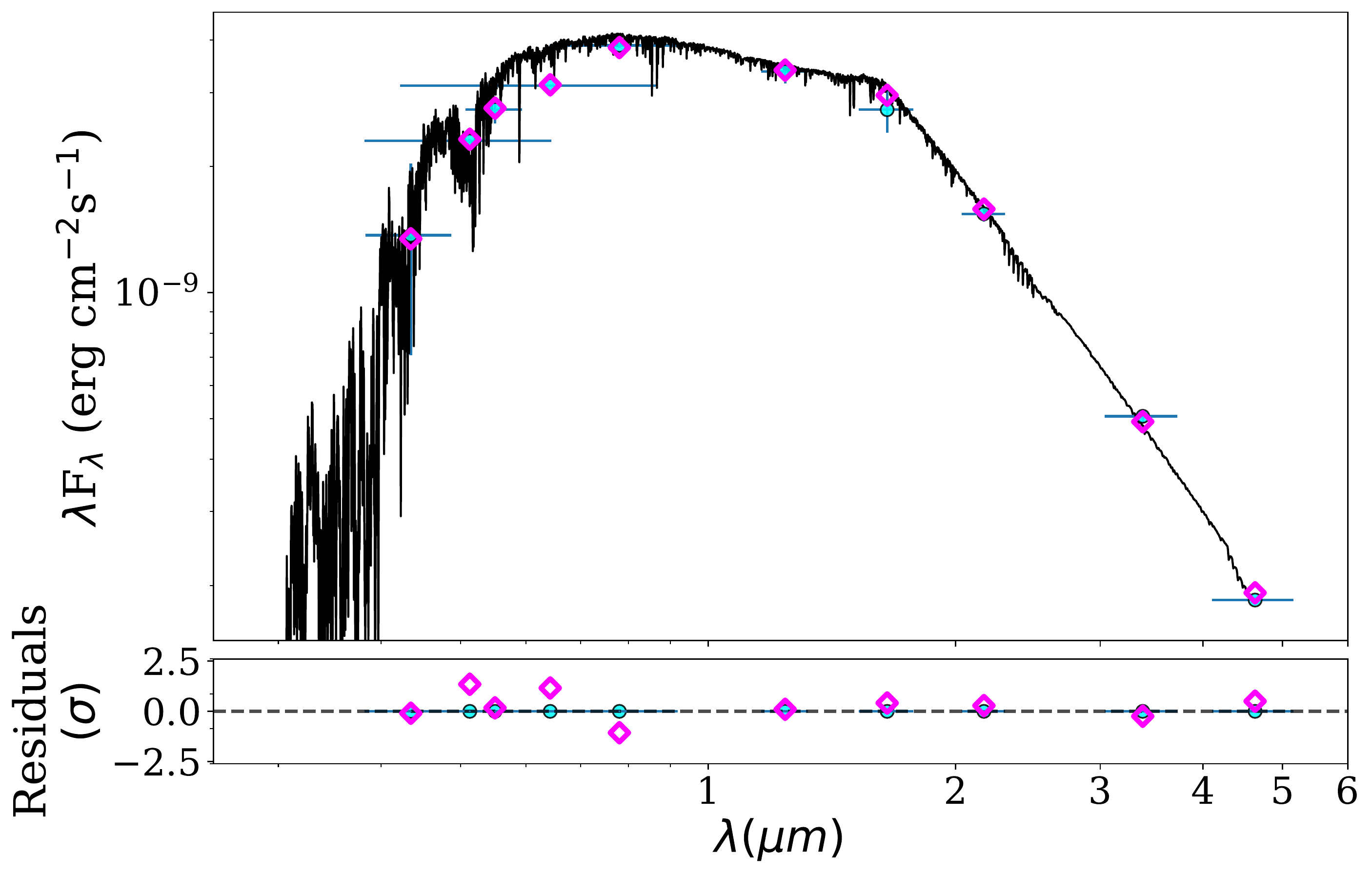}
    \caption{The SED of \hd\ and the model calculated from the model grid with the highest probability. Magenta points are the synthetic photometry and the blue points the observed photometry where the vertical errors mark the $1~\sigma$ uncertainties and the horizontal bars the effective width of the pass-bands. The residuals in the lower panel are normalised to the errors of the photometry.}
    \label{fig:sed}
\end{figure}

As an extra check to our stellar mass and radius determination, we also perform a Spectral Energy Distribution (SED) modelling using the software  \href{https://github.com/jvines/astroARIADNE}{{\tt{ARIADNE}\,\faGithub}}  
\citep[][]{2020MNRAS.498.3115A}.  Grids of four stellar atmospheric models,    {\tt {Phoenix~v2}} \citep{2013A&A...553A...6H}, {\tt {BtSettl}} \citep{2012RSPTA.370.2765A}, \citet{Castelli2004}, and \citet{Kurucz-93} were interpolated with priors for  $T_\mathrm{eff}$, $\log g_\star$,  [Fe/H]   from {\tt{SpecMatch-Emp}} (Table~\ref{tab:stellarparams}). Stellar radius, distance, and extinction  ($A_V$) were treated as free parameters. We used broad band photometry from  2MASS J, H, and K,  {\it WISE} W1 and W2, the  {\it Johnson} B and V magnitudes from APASS, and {\it Gaia} G, G$_{\rm BP}$, and G$_{\rm RP}$ magnitudes  and parallax from eDR-3. An upper limit to the extinction was taken from the maximum line-of-sight value  from the dust maps of \citet{1998ApJ...500..525S}. Bayesian Model Averaging was used to compute the final stellar radius which was found to be $0.693^{+0.019}_{-0.028}$~$R_\odot$.  {\tt{ARIADNE}} also computed the  stellar mass using MIST \citep{2016ApJ...823..102C} stellar evolution tracks to  $0.727^{+0.022}_{-0.030}$~$M_\odot$. These results agrees within one sigma with the above derived mass and radius from \texttt{PARAM\,1.3}. Figure~\ref{fig:sed} shows the SED fit obtained with {\tt{ARIADNE}}, and the corresponding results are summarised in Table~\ref{tab:stellarparams}.

{
As a further check on the reliability of our stellar parameters, we also performed some basic checks within our exoplanet analyses (see Sect.~\ref{sec:dataanalysis}). 
We compare the planetary surface gravity obtained from the derived planet masses and radii (that depends on the derived stellar mass and radius) with the value obtained from the scaled parameters \citep[that do not depend on the stellar mass and radius, see][]{Sotuhworth2007}. Both values for both planets are in agreement within $<0.1 \sigma$ (see Table~\ref{tab:derived}).
These results suggest that the derived stellar parameters in this section are reliable.
}

\subsection{Stellar rotation period}
\label{sec:perwasp}

We note that \tess\ photometry shows flux modulation in both \hd's light curves (see Sect.~\ref{sec:tess} and Fig.~\ref{fig:lcs}). These are likely caused by active regions on the stellar surface. Fortunately, they can be a proxy to estimate the stellar rotation period. A Fourier transform of Sector 8 and 34 light curves shows peaks at $\sim 12$ and $\sim 6$ days{, being} the later likely the first harmonic caused by a complex distribution of spots in the stellar surface. Given that the \tess\ observations cover only $\sim 2$ rotational period of the star, we decide to use ground-base photometry that expands for a longer time window to estimate a more precise stellar rotational period.

\begin{center}
\begin{figure}
\includegraphics[width=0.49\textwidth]{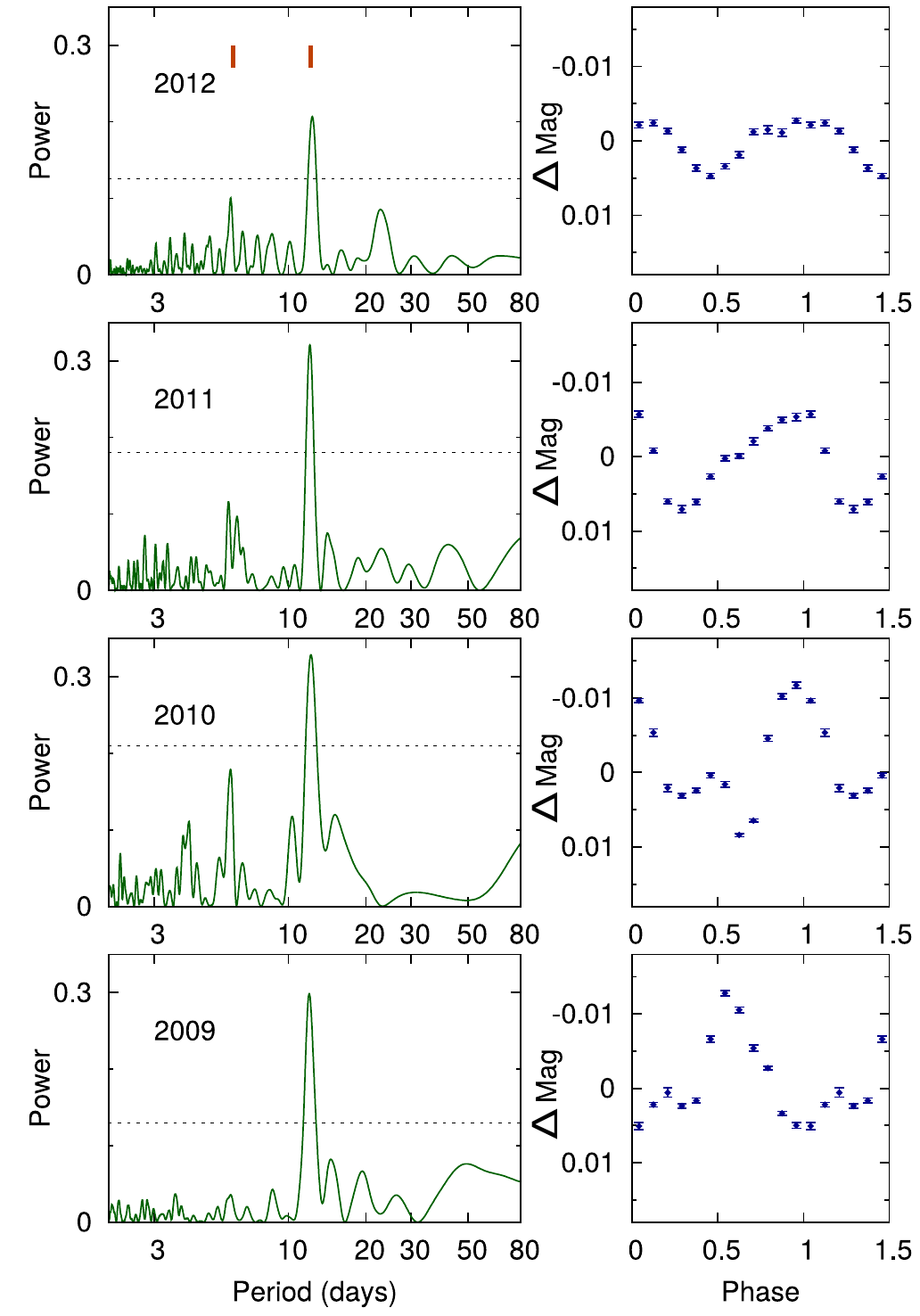}
  \caption{Periodograms of the WASP-South data for \hd. Red ticks mark the 12.2-d periodicity and its first harmonic. The dotted line is the estimated 1\%-likelihood false-alarm level. The right-hand panels show the data binned and folded on the 12-d period.}
\label{fig:wasp}
\end{figure}
\end{center}

\hd\ was observed with the WASP-South \citep{2006PASP..118.1407P} camera array over 100-night spans in four consecutive years from 2009 to 2012, accumulating 17\,000 photometric observations. The 200-mm, f/1.8 lenses were backed by a 2048x2048 CCDs observing with a 400--700 nm passband. 
\hd\ is 2 magnitudes brighter than any other star in the 48-arcsec photometric aperture.  We searched the data for rotational modulations using the methods from \citet{2011PASP..123..547M}, and found a clear and persistent 12-d periodicity with a false-alarm likelihood below 0.1\%.  The amplitude ranges from 4 to 7 mmag while the period is 12.2 $\pm$ 0.2 d, where the error makes allowance for phase shifts caused 
by changing star-spot patterns. Figure~\ref{fig:wasp} shows a visualisation of our analysis.
We note that this period is consistent with the period of \jPGP\,d recovered in our multidimensional GP analysis (for more details see Sec.~\ref{sec:multigp}).

\subsection{Stellar age}
\label{sec:stellarage}

Given \hd\ spectral type and well-constrained relatively short rotational period, we expect a relative young age \citep[see][for more details]{Barnes2003}.
We found that the derived age for this star in the literature provides poorly constrained ages \citep[e.g.,][ estimated a stellar age of $5.12 \pm 4.56$\,Gyr]{Delgado2019}. We therefore perform some analyses to re-estimate \hd's age.

{
We first use \href{https://github.com/RuthAngus/stardate}{\texttt{stardate}\,\faGithub}  \citep{stardate,isochrones} to infer \hd's age using gyrochronology. We input the \texttt{SpecMatch-Emp} spectroscopic values, together with the photometry band values and parallax given in Table~\ref{tab:parstellar} to \texttt{stardate}. For the rotational period of the star, we use the rotational period of the star of \jPGP\,d derived in our multidimensional GP analysis of the spectroscopic time-series (see Sect.~\ref{sec:dataanalysis}).
We ran 100\,000 iterations and we discarded the first 10\,000 to create the distributions from which we infer our parameters. \texttt{stardate} gives an age of \hd\ of $750 \pm 20$ Myr. This analysis puts \hd\ in the young star regime. However, we note that the error bars come only from the built-in parameter sampling included in \texttt{stardate} and they are likely underestimated. We therefore perform additional estimations to provide a more conservative stellar age. 

We then use the time-\logr relation of \citet{Mamajek2008} to estimate \hd's age.
With a \logr$= -4.465 \pm 0.015$, \hd's is consistent with a star with an age of $480 \pm 190$~Myr. The error bars include the uncertainties of \logr\ and the 30\% rms on the \citet{Mamajek2008}'s fit. In the remainder of this paper, we will assume $480 \pm 190$~Myr as the stellar age.
We note that this stellar age is consistent with the recently reported values by \citet[][]{Zhang2021}.
}



\section{Exoplanet data analysis}
\label{sec:dataanalysis}

\subsection{Transit analysis}
\label{sec:transitanalysis}

We first perform a transit analysis in order to obtain planet ephemerides  to use for a preliminary RV data analysis in Sect.~\ref{sec:multigp}, as well as to check for uniform transit depths in all bands for both planets. 
We use the code \pyaneti\ \citep{pyaneti,pyaneti2} to model the flattened \tess\ (see Sect.~\ref{sec:tess}), \cheops\ (see Sect.\ref{sec:cheops}), and ground-base transits (see Sect.~\ref{sec:fut}). 
To speed-up the analysis, we just model data chunks of maximum 3.5 hours either side of each transit mid-time.

We sample for the stellar density, $\rho_\star$, and we recover the scaled semi-major axis ($a/R_\star$) for both planets using Kepler's third law \citep[see e.g.,][]{Winn2010}. 
For the limb darkening model we use the quadratic limb darkening approach described in \citet{Mandel2002} with the $q_1$ and $q_2$ parametrisation given by \citet{Kipping2013}. 
We sample for an independent scaled planet radius ($r_{\rm p} \equiv R_{\rm p}/R_\star$) for each planet in each band.
We set uniform priors for all the parameters and we assume circular orbits for both planets. 
We sample the parameter space with 250 walkers using the Markov chain Monte Carlo (MCMC) ensemble sampler algorithm implemented in \pyaneti\ \citep{pyaneti,emcee}. 
We created the posterior distributions with the last 5000 iterations of converged chains. We thinned our chains with a factor of 10 giving a distribution of 125\,000 independent points for each sampled parameter.

\newcommand{\Tzerobtr}[1][BTJD]   {$1517.69016 \pm 0.00063$~#1} 
\newcommand{\Pbtr}[1][days]   {$6.3980422 \pm 0.0000070$~#1} 
\newcommand{\rrbtess}{$0.03984 _{ - 0.00064 } ^ { + 0.00079 }$} 
\newcommand{\rrbngts}{$0.0404 _{ - 0.0015 } ^ { + 0.0015 }$} 
\newcommand{\rrblcozs}{$0.0412 _{ - 0.0016 } ^ { + 0.0016 }$} 
\newcommand{\rrblcoB}{$0.0394 _{ - 0.0019 } ^ { + 0.0017 }$} 
\newcommand{\rrbpest}{$0.0494 _{ - 0.0032 } ^ { + 0.0031 }$} 
\newcommand{\rrbcheops}{$0.03962 _{ - 0.00091 } ^ { + 0.00101 }$} 
\newcommand{\Tzeroctr}[1][BTJD]   {$1232.1682 \pm 0.0025$~#1} 
\newcommand{\Pctr}[1][days]   {$18.87981 \pm 0.00090 $~#1} 
\newcommand{\rrctess}{$0.03325 _{ - 0.00073 } ^ { + 0.00075 }$} 
\newcommand{\rrclcozs}{$0.0344 _{ - 0.0025 } ^ { + 0.0023 }$} 
\newcommand{\dentrtr}[1][${\rm g\,cm^{-3}}$]   {$3.05 _{ - 0.69 } ^ { + 0.51 }$~#1} %

\begin{figure*}
    \centering
    \includegraphics[width=\textwidth]{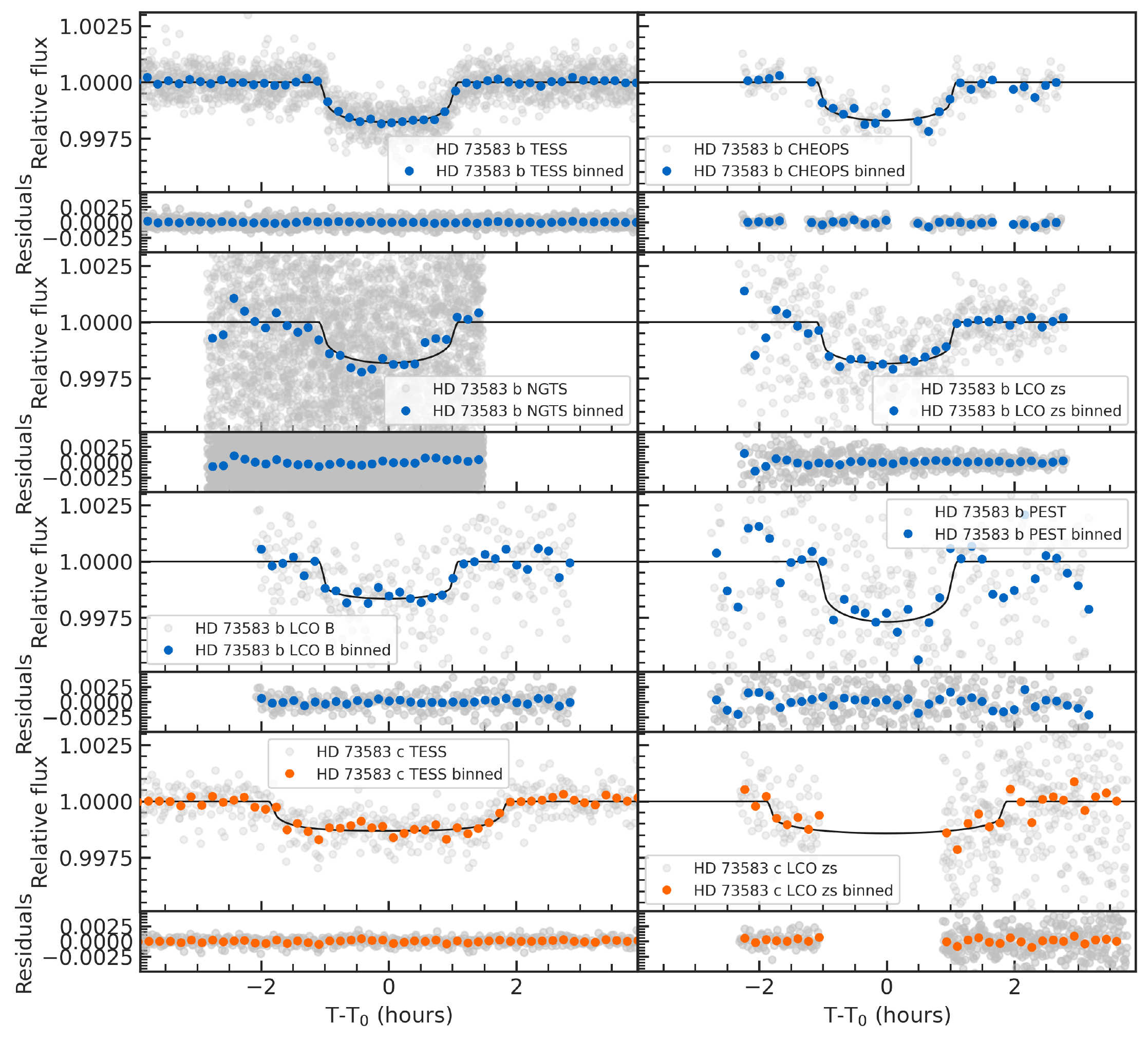}\\
    \caption{Phase-folded light curves of \hdb\ (Panels with blue circles) and \hdc\ (Panels with orange circles) for different bands. Nominal observations are shown in light grey. Solid colour circles represent 10-min binned data.
    Transit models are shown with a solid black line.}
    \label{fig:trfolded}
\end{figure*}

\begin{figure}
    \centering
    \includegraphics[width=0.48\textwidth]{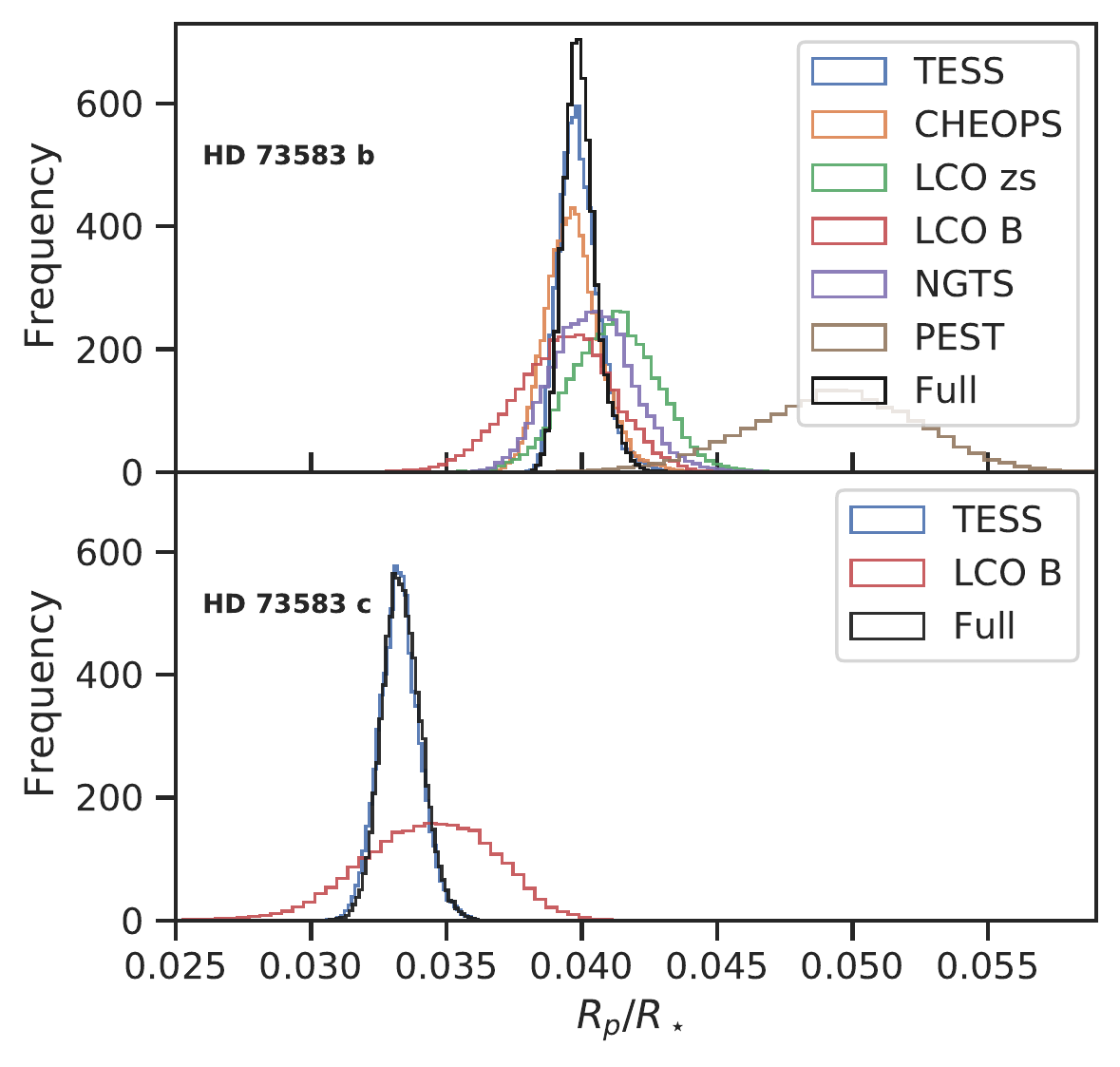}
    \caption{Posterior distributions for the multi-radius (colourful posteriors) and single-radius (black posteriors) transit modellings, for planet b (top panel) and c (bottom panel). Both panels are shown with the same x- and y-scale to facilitate comparison.}
    \label{fig:posteriorradii}
\end{figure}

Figure~\ref{fig:trfolded} shows the phase-folded transits for \hdbc\ for all the bands they were observed. Figure~\ref{fig:posteriorradii} shows the posterior distribution for each sampled $r_{\rm p}$ for both planets.
We can see that for \hdb, the posteriors for $r_{\rm p}$ for all bands overlap between them, except for the PEST light curve that is consistent with the rest of estimation just within 3 sigma.
The corresponding scaled planet radii are $r_{\rm p, \emph{TESS}} = $\rrbtess, $r_{\rm p, \emph{CHEOPS}} =\, $\rrbcheops, $r_{\rm p,NGTS} =\, $\rrbngts, $r_{\rm p, LCO-B} =\, $\rrblcoB, $r_{\rm p, LCO-zs} =\, $\rrblcozs, and $r_{\rm p, PEST} =\, $\rrbpest.
\hdc\ was only observed with \tess\ and LCO-zs; the scaled planet radii are consistent in these two bands ($r_{\rm p, \emph{TESS}} =\, $\rrctess, and $r_{\rm p, LCO-zs} =\, $\rrclcozs, see also the posteriors in Fig.~\ref{fig:posteriorradii}). { We also note that the recovered stellar density from this analysis is \dentrtr, consistent with the value obtained in the stellar analysis presented in Sect.~\ref{sec:stellarmassradius}.}

We then repeated the analysis, but this time sampling for a single $r_{\rm p,full}$ for each planet, i.e., assuming that, for a given planet, all the transits have the same depth in all bands.
Figure~\ref{fig:posteriorradii} shows the posterior distribution obtained in this analysis for \hdbc.
We can see that for \hdb\ the combination of all transits in all bands provides a better constrain on the scaled planetary radius of $r_{\rm p,full} = 0.0399 _{ - 0.00054 } ^ { + 0.00064 }$. This shows the advantages of performing space- and ground-base transit follow-up of \tess\ planets to improve the measurement of planetary radii.
For \hdc\ we can see that the posterior of $r_{\rm p,full}$ is practically identical to the posterior of $r_{\rm p, \emph{TESS}}$, this is expected given that the ground-base observations of \hdc\ observed only partial transits.

We also note that all transit data improve the constrain on the ephemerides that have a direct impact to plan future follow-up observations. 
We obtain a time of mid-transit and orbital period of \Tzerobtr\ and \Pbtr\ for \hdb, 
and \Tzeroctr\ and \Pctr\ for \hdc, respectively. We use these values to perform our spectroscopic time-series modelling in Sect.~\ref{sec:multigp}.
In Sect.~\ref{sec:joint} we perform a joint analysis with transits and spectroscopic time-series.



\subsection{Multidimensional GP approach}
\label{sec:multigp}

We perform a multidimensional GP (hereafter multi-GP) approach to characterise the stellar and planetary signals in our RV time-series \citep[see][for more details]{Rajpaul2015}.
We create a 2-dimensional GP model described as
\begin{equation}
    \begin{aligned}
    RV & = & A_{\rm RV} G(t) & + B_{\rm RV} \dot{G}(t), \\
    S_{\rm HK} & = & A_{\rm S} G(t),
    \label{eq:3mgp}
\end{aligned}
\end{equation}
\noindent where $G(t)$ {is a latent (unobserved) variable, which can be loosely interpreted as representing the projected area of the} visible stellar disc that is covered in active regions at a given time. The amplitudes $A_{\rm RV}$, $B_{\rm RV}$, and $A_{\rm S}$ are free parameters which relate the individual time-series with $G(t)$. To constrain the stellar signal in our data we use \sshk\ that has been proven to be a good tracer of the area covered for active regions on the stellar surface \citep[see e.g.,][]{Isaacson2010,Thompson2017}.
It is also worth to mention that the \ion{Calcium}{II} H~\&~K activity indicators (\sshk\ and \logr)  have been proved to constrain the $G(t)$ function for active G and K-type stars in previous multi-GP analyses \citep[e.g.][]{Barragan2019,Georgieva2021}. 

We perform a multi-GP regression on the HARPS, HIRES, PFS, and CORALIE RV and \sshk\ time-series. 
We created our covariance matrix using the Quasi-Periodic kernel
\begin{equation}
    \gamma(t_i,t_j) = \exp 
    \left[
    - \frac{\sin^2[\pi(t_i - t_j)/P_{\rm GP}]}{2 \lambda_{\rm P}^2}
    - \frac{(t_i - t_j)^2}{2\lambda_{\rm e}^2}
    \right],
    \label{eq:gamma}
\end{equation}
and its derivatives \citep[see][for more details]{pyaneti2,Rajpaul2015}.
In equation~\eqref{eq:gamma} \pgp\ is the GP characteristic period, \lbp\ the inverse of the harmonic complexity, and \lbe\ is the long term evolution timescale. 

We performed a multi-GP regression using \pyaneti\ \citep[as described in][]{pyaneti2}.
We created our RV residual vector by {modelling} an offset for each spectrograph and two Keplerian signals with the ephemeris given in our transit analysis and assuming circular orbits.
For the \sshk\ time-series we account for a different offset for each instrument.
We note that we are using four different spectrographs that observe in similar wavelength ranges and we expect the spot contrast to be similar. 
For this reason we consider that there is no chromatic variation between them and we assume that the stellar activity can be described with the same underlying function for all four instruments.

We perform an MCMC analysis using Gaussian priors on the planet ephemerides given in Sect.~\ref{sec:transitanalysis}. For the remainder sampled parameters we use uniform priors. 
We note that we did not train our GP hyper-parameters using the \tess\ or WASP-South light curves \citep[see][for more details on training GPs for RV modelling]{Haywood2014}. The active regions on the stellar surface may be different and therefore the stellar signal may be described with a GP with a different set of hyper-parameters \citep[see e.g.,][]{Barragan2021b}. We therefore use uniform priors to sample for the multi-GP hyper-parameters (we chose an uniform prior around 12 days for the period based on the analyses presented in Sect.~\ref{sec:perwasp}).
We sample the parameter space with 250 independent Markov chains and we use the last 5000 converged chains and a thin factor of 10 to create posterior distributions with 125\,000 independent samples for each parameter. 
We obtained uni-modal posterior distributions for all the sampled parameters. 
This analysis provides a detection of two Keplerian signals that match the transiting exoplanet ephemeris. \hdbc\ induce a Doppler wobble with semi-amplitudes of $4.3_{-1.15}^{+1.43}$\,\ms and $2.74_{-0.56}^{+0.57}$\,\ms on its host star, respectively.
In Sect.~\ref{sec:stellarsignal} we make a further discussion on the inferred stellar and planetary signals.

{
It is worth to note that the precision of the recovered Doppler signal of \hdc\ is better than for \hdb. This is a consequence of the orbital period of \hdb\ ($\sim 6.4$\,d) being close to the first harmonic of the rotational period of the star ($\sim 6$\,d). This imperils the precision with which \hdb\ Keplerian signal can be recovered.}

\subsection{Joint analysis}
\label{sec:joint}

Following our analyses described in Sections~\ref{sec:transitanalysis} and \ref{sec:multigp}, we proceed to perform a final analysis combining our transit and spectroscopic time-series modelling. 
This analysis combines the same assumptions described in Sections~\ref{sec:transitanalysis} and \ref{sec:multigp} with a few differences. Now we allow for eccentric orbits for both planetary orbits; we set a Gaussian prior on the stellar density based on our results in Sect.~\ref{sec:stellar}, and we fit for a common scaled planet radius for all the bands for each planet.
All sampled parameters and priors used for this analysis can be found in Table~\ref{tab:pars}. We note that fourteen of the sampled parameters are jitter terms per instrument that we added in our likelihood to penalise for imperfections in our model \citep[see][]{pyaneti2}.
We sample the parameter space with 500 independent Markov chains.
We create posterior distributions with 250\,000 sampled points for each parameter using the last 25\,000 converged chains and a thin factor of 50.
Given the high dimensionality of the parameter space (55 parameters), we ran this setup 10 times. We found that the code arrives to consistent parameter solutions in every independent run. This give us confidence that the derived parameters are reliable.  

\begin{figure*}
    \centering
    \includegraphics[width=0.94\textwidth]{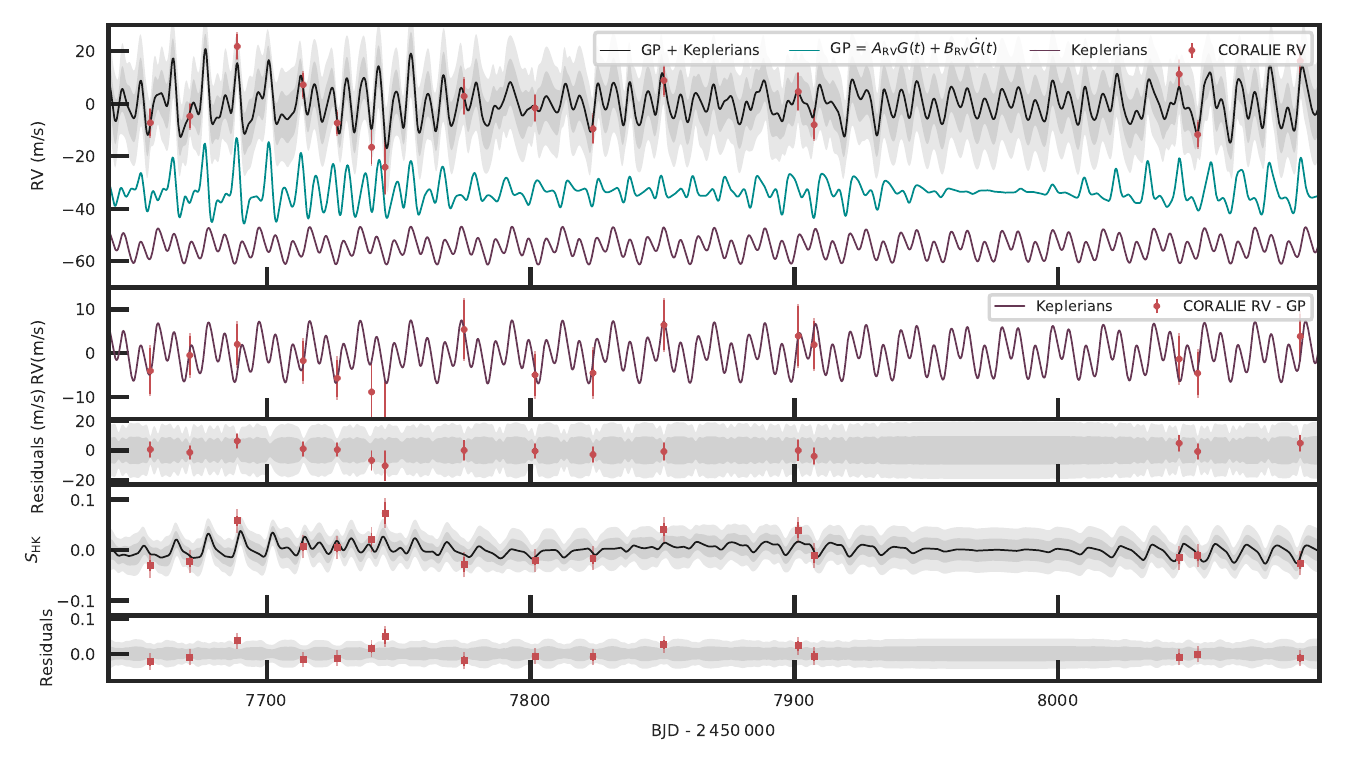}
    \includegraphics[width=0.94\textwidth]{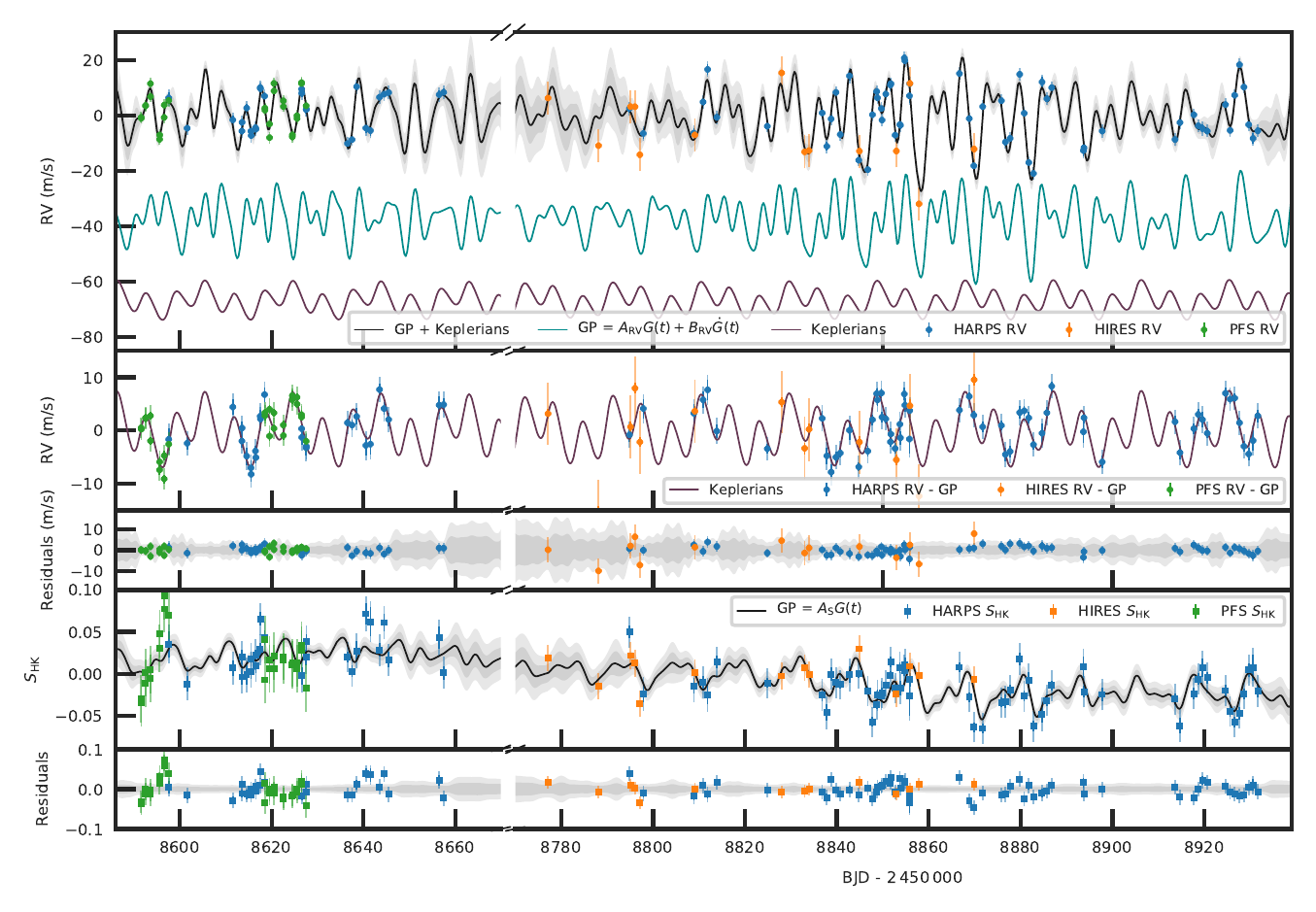}
    \caption{Radial velocity and \sshk\ time-series after been corrected by inferred offsets. Each plot shows (from top to bottom): RV data together with full, stellar, and planetary signal inferred models; RV data with stellar signal model subtracted; RV residuals; \sshk\ data together with inferred stellar model, and \sshk\ residuals.
    Upper plot shows CORALIE (red) observations. Bottom plot displays HARPS (blue), HIRES (orange), and PFS (green) data.  
    Measurements are shown with filled symbols with error bars with a semi-transparent error bar extension accounting for the inferred jitter. 
    The solid (black) lines show the inferred full model coming from our multi-GP, light grey shaded areas showing the one and two sigma credible intervals of the corresponding GP model.
    For the RV time-series we also show the inferred stellar (cyan line) and planetary (dark purple line) recovered signals with an offset for better clarity.
    We note that in the bottom plot there is a gap between 8670 and 8770 BJD - 2\,450\,000 where there were no measurements.}
    \label{fig:timeseries}
\end{figure*}

\begin{figure*}
    \centering
    \includegraphics[width=0.49\textwidth]{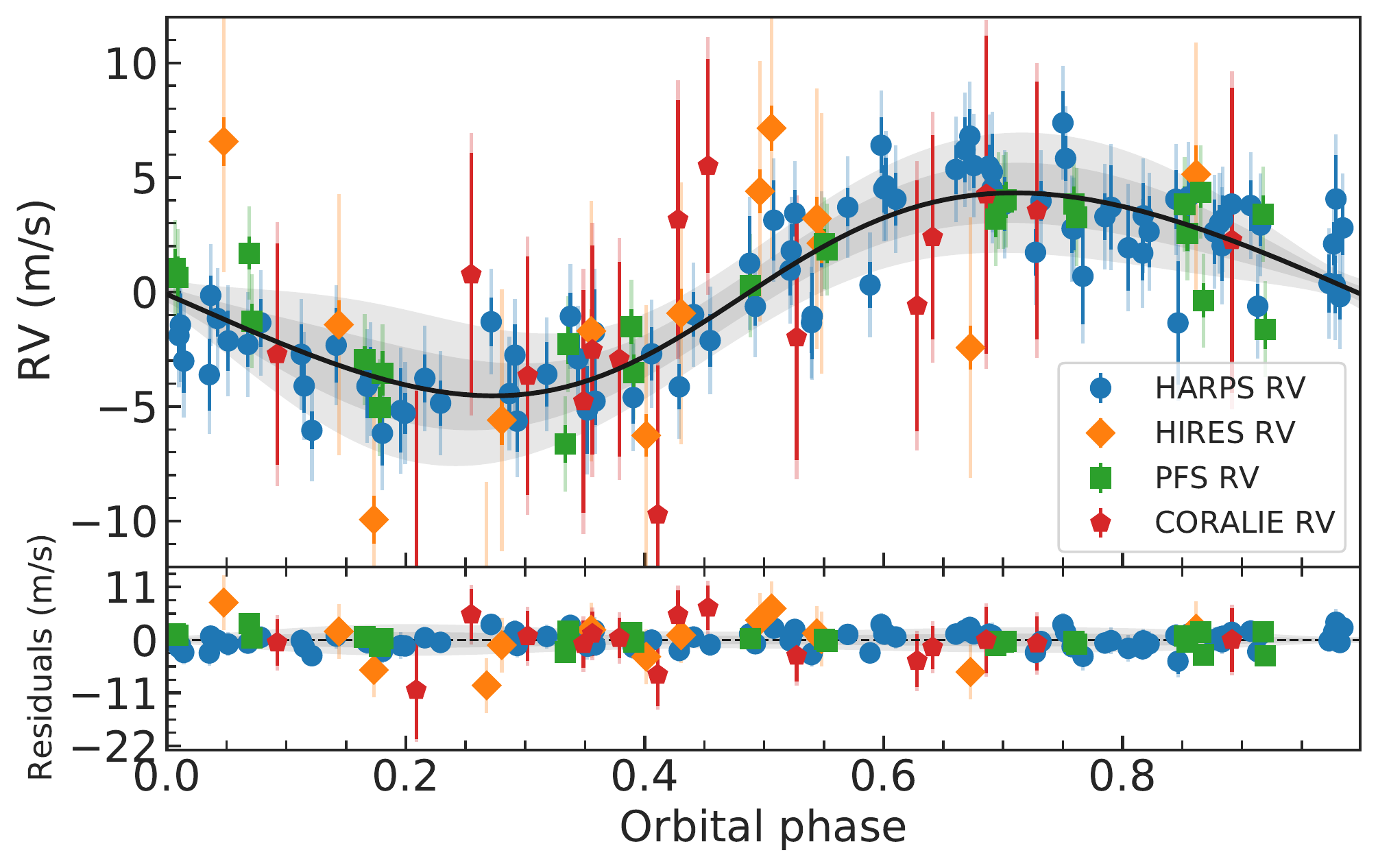}
    \includegraphics[width=0.49\textwidth]{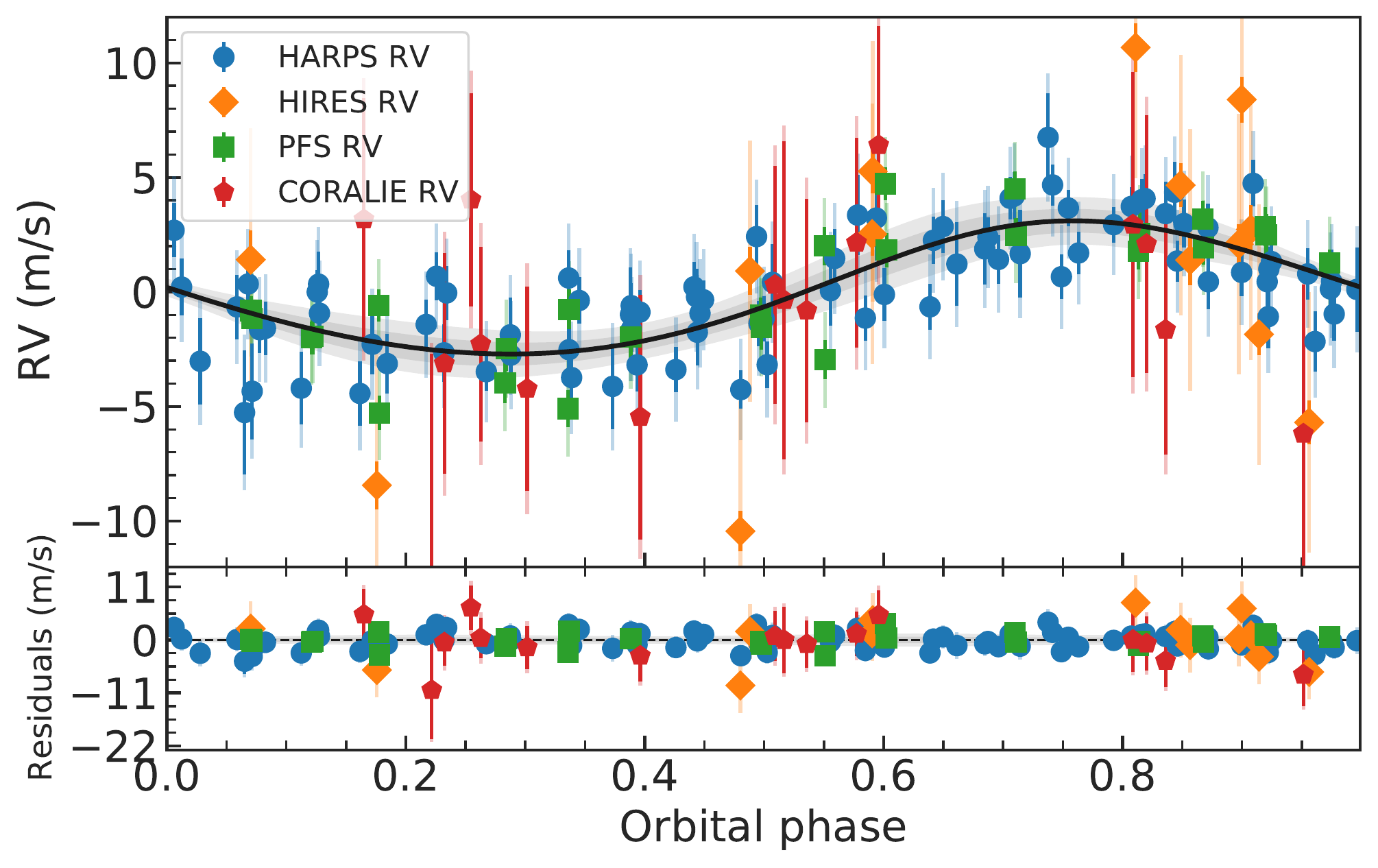}
    \caption{Phase-folded RV signals for \hdb\ (left) and \hdc\ (right) following the subtraction of the systemic velocities, stellar signal, and other planets. HARPS (blue circles), HIRES (orange diamonds), PFS (green squares) and CORALIE (red pentagons) RV observations are shown.
    RV models are shown (solid black line) with 1 and 2 sigma credible intervals (shaded areas).
    In all the plots the nominal error bars are in solid colour, and the error bars taking into account the jitter are semitransparent.}
    \label{fig:rvfolded}
\end{figure*}

Table~\ref{tab:pars} shows the inferred values for all the sampled parameters. They are given as the median and 68.3\% region of the credible interval from the posterior distribution of each parameter. In Figure~\ref{fig:correlations} we show the posterior and correlation plots for some of the sampled parameter.
Table~\ref{tab:derived} shows the derived planetary, orbital, and stellar parameters.
\hdbc\ are detected in the RV time-series with Doppler semi-amplitudes of \kb\ and \kc, respectively. 
Figure~\ref{fig:timeseries} shows the RV and \sshk\ time-series together with the inferred models. Figure~\ref{fig:rvfolded} shows the phase-folded RV curve for each planetary induced signal.

\begin{table*}
\begin{center}
  \caption{Model parameters and priors for joint fit \label{tab:pars}}  
  \begin{tabular}{lcc}
  \hline
  \hline
  \noalign{\smallskip}
  Parameter & Prior$^{(a)}$ & Final value$^{(b)}$ \\
  \noalign{\smallskip}
  \hline
  \noalign{\smallskip}
  \multicolumn{3}{l}{\emph{\bf \hdb's parameters }} \\
  \noalign{\smallskip}
    Orbital period $P_{\mathrm{orb}}$ (days)  & $\mathcal{U}[ 6.3973 , 6.3982]$ &\Pb[] \\
    Transit epoch $T_0$ (BJD$_\mathrm{TDB}-$2\,450\,000)  & $\mathcal{U}[8517.5460 , 8517.8040]$ & \Tzerob[]  \\  
    Scaled planet radius $R_\mathrm{p}/R_{\star}$  &$\mathcal{U}[0.0,0.05]$ & \rrb[]  \\
    Impact parameter, $b$ & $\mathcal{U}[0,1]$ & \bb[] \\
    $\sqrt{e} \sin \omega_\star$  & $\mathcal{U}[-1,1]$ & \esinb[]  \\
    $\sqrt{e} \cos \omega_\star$ &  $\mathcal{U}[-1,1]$ & \ecosb[]  \\
    Doppler semi-amplitude variation $K$ (m s$^{-1}$) & $\mathcal{U}[0,50]$ & \kb[] \\
    \multicolumn{3}{l}{\emph{ \bf \hdc's parameters}} \\
    Orbital period $P_{\mathrm{orb}}$ (days)  & $\mathcal{U}[18.78 , 18.98]$ &\Pc[] \\
    Transit epoch $T_0$ (BJD$_\mathrm{TDB}-$2\,450\,000)  & $\mathcal{U}[9232.06 , 9232.26 ]$ & \Tzeroc[]  \\  
    Scaled planet radius $R_\mathrm{p}/R_{\star}$  &$\mathcal{U}[0.0,0.05]$ & \rrc[]  \\
    Impact parameter, $b$ & $\mathcal{U}[0,1]$ & \bc[] \\
    $\sqrt{e} \sin \omega_\star$  & $\mathcal{U}[-1,1]$ & \esinc[]  \\
    $\sqrt{e} \cos \omega_\star$ &  $\mathcal{U}[-1,1]$ & \ecosc[]   \\
    Doppler semi-amplitude variation $K$ (m s$^{-1}$) & $\mathcal{U}[0,50]$ & \kc[] \\
    \multicolumn{3}{l}{\emph{ \bf GP hyper-parameters}} \\
   GP Period $P_{\rm GP}$ (days) &  $\mathcal{U}[10,14]$ & \jPGP[] \\
    $\lambda_{\rm p}$ &  $\mathcal{U}[0.1,5]$ &  \jlambdap[] \\
    $\lambda_{\rm e}$ (days) &  $\mathcal{U}[1,500]$ &  \jlambdae[] \\
    $A_{\rm RV}$ (\ms)  &  $\mathcal{U}[0,100]$ & \jAzero \\
    $B_{\rm RV}$ (\ms\,d) &  $\mathcal{U}[-1000,1000]$ & \jAone \\
    $A_{\rm S}$  &  $\mathcal{U}[0,1]$ & \jAtwo \\
    \multicolumn{3}{l}{\emph{ \bf Other parameters}} \\
    Stellar density $\rho_\star$ (\gcm) & $\mathcal{N}$[3.48,0.35] & \denstrb[] \\ 
    \tess\ Parameterised limb-darkening coefficient $q_1$  &$\mathcal{U}[0,1]$ & \qonetess \\ 
    \tess\ Parameterised limb-darkening coefficient $q_2$  &$\mathcal{U}[0,1]$ & \qtwotess \\ 
    CHEOPS Parameterised limb-darkening coefficient $q_1$  &$\mathcal{U}[0,1]$ & \qonecheops \\ 
    CHEOPS Parameterised limb-darkening coefficient $q_2$  &$\mathcal{U}[0,1]$ & \qtwocheops \\ 
    NGTS Parameterised limb-darkening coefficient $q_1$  &$\mathcal{U}[0,1]$ & \qonengts \\ 
    NGTS Parameterised limb-darkening coefficient $q_2$  &$\mathcal{U}[0,1]$ & \qtwongts \\ 
    LCO-zs Parameterised limb-darkening coefficient $q_1$  &$\mathcal{U}[0,1]$ & \qonelcozs \\ 
    LCO-zs Parameterised limb-darkening coefficient $q_2$  &$\mathcal{U}[0,1]$ & \qtwolcozs \\ 
    LCO-B Parameterised limb-darkening coefficient $q_1$  &$\mathcal{U}[0,1]$ & \qonelcoB \\ 
    LCO-B Parameterised limb-darkening coefficient $q_2$  &$\mathcal{U}[0,1]$ & \qtwolcoB \\ 
    PEST Parameterised limb-darkening coefficient $q_1$  &$\mathcal{U}[0,1]$ & \qonepest \\ 
    PEST Parameterised limb-darkening coefficient $q_2$  &$\mathcal{U}[0,1]$ & \qtwopest \\ 
    Offset HARPS RV (\kms) & $\mathcal{U}[ 20.2311 , 21.2728]$ & \HARPSRV[] \\
    Offset PFS RV (\kms) & $\mathcal{U}[ -0.5253 , 0.5219]$ & \PFSRV[]  \\
    Offset HIRES RV (\kms) & $\mathcal{U}[ -0.5253 , 0.5219]$ & \HIRESRV[]  \\
    Offset CORALIE RV (\kms) & $\mathcal{U}[ 20.2129 , 21.2589]$ & \CORALIERV[]  \\
    Offset HARPS $S_{\rm HK}$ & $\mathcal{U}[0.3141 , 1.4507 ]$ & \HARPSSHK[]  \\
    Offset PFS $S_{\rm HK}$ & $\mathcal{U}[-0.0856 , 1.0419 ]$ &  \PFSSHK[]  \\
    Offset HIRES $S_{\rm HK}$ & $\mathcal{U}[0.1008 , 1.1666 ]$ & \HIRESSHK[]   \\
    Offset CORALIE $S_{\rm HK}$ & $\mathcal{U}[ 0.1956 , 1.2995 ]$ & \CORALIESHK[]  \\
    Jitter term $\sigma_{\rm RV, HARPS}$ (\ms) & $\mathcal{J}[1,100]$ & \jHARPSRV[] \\
    Jitter term $\sigma_{\rm RV, PFS}$ (\ms) & $\mathcal{J}[1,100]$ &  \jPFSRV[] \\
    Jitter term $\sigma_{\rm RV, HIRES}$ (\ms) & $\mathcal{J}[1,100]$ & \jHIRESRV[] \\
    Jitter term $\sigma_{\rm RV, CORALIE}$ (\ms) & $\mathcal{J}[1,100]$ & \jCORALIERV[] \\
    Jitter term $\sigma_{\rm S_{HK}, HARPS} (\times 10^3)$  & $\mathcal{J}[1,100]$ & \jHARPSSHK[] \\
    Jitter term $\sigma_{\rm S_{HK}, PFS} (\times 10^3)$  & $\mathcal{J}[1,100]$ & \jPFSSHK[] \\
    Jitter term $\sigma_{\rm S_{HK}, HIRES} (\times 10^3)$  & $\mathcal{J}[1,100]$ & \jHIRESRV[] \\
    Jitter term $\sigma_{\rm S_{HK}, CORALIE} (\times 10^3)$ & $\mathcal{J}[1,100]$ & \jCORALIESHK[] \\
    \tess\ jitter term $\sigma_{\tess}$ ($\times 10^{-6}$) & $\mathcal{J}[0,1 \times10^{3}]$ & \jtrtess \\
    CHEOPS jitter term $\sigma_{CHEOPS}$ ($\times 10^{-6}$) & $\mathcal{J}[0,1 \times10^{3}]$ & \jtrcheops \\
    NGTS jitter term $\sigma_{NGTS}$ ($\times 10^{-6}$) & $\mathcal{J}[0,1 \times10^{3}]$ & \jtrngts \\
    LCO-zs jitter term $\sigma_{LCO-zs}$ ($\times 10^{-6}$) & $\mathcal{J}[0,1 \times10^{3}]$ & \jtrlcozs \\
    LCO-B jitter term $\sigma_{LCO-B}$ ($\times 10^{-6}$) & $\mathcal{J}[0,1 \times10^{3}]$ & \jtrlcoB \\
    PEST jitter term $\sigma_{PEST}$ ($\times 10^{-6}$) & $\mathcal{J}[0,1 \times10^{3}]$ & \jtrpest \\
    \noalign{\smallskip}
    \hline
\multicolumn{3}{l}{\footnotesize $^a$ $\mathcal{U}[a,b]$ refers to uniform priors between $a$ and $b$, $\mathcal{N}[a,b]$ to Gaussian priors with mean $a$ and standard deviation $b$, and $\mathcal{J}[a,b]$ to the }\\
\multicolumn{3}{l}{ modified Jeffrey's prior as defined by \citet[eq.~16]{Gregory2005}.}\\
\multicolumn{3}{l}{\footnotesize $^b$ Inferred parameters and errors are defined as the median and 68.3\% credible interval of the posterior distribution.}
  \end{tabular}
\end{center}
\end{table*}

\section{Discussion and conclusions}
\label{sec:discusion}

\subsection{Stellar signal characterisation}
\label{sec:stellarsignal}

The GP hyper-parameters are well constrained with the values \pgp$=12.024_{-0.087}^{+0.095}$\,d, \lbe$=32_{-6}^{+7}$\,d, and \lbp$=0.432_{-0.046}^{+0.050}$.  We can argue that \pgp\ describes the star's rotation period, given that is in agreement with the results obtained with WASP-South photometry (Sect.~\ref{sec:perwasp}).
The recovered \lbe\ is significantly larger than the recovered \pgp. This suggests that the same active regions on the stellar surface are present for almost three stellar rotations. This also indicates that the QP kernel is a good choice to describe the scales of the stellar signal in our data \citep[see][for a discussion on this]{Rajpaul2015}.
Finally, the recovered \lbp\ implies a relatively high harmonic complexity for the underlying process describing the stellar signal. This suggests that there are diverse groups of active regions on the stellar surface, leading to complex patterns in our spectroscopic time-series.

Figure~\ref{fig:timeseries} shows the inferred model for the stellar signal in the RV and \sshk\ time-series. Such curves are consistent with a high harmonic complexity scenario having several ``beat'' patterns within each period. 
However, the process describing the RV stellar signal has an apparent higher harmonic complexity than the \sshk\ one. 
This can be explained by the sensitivity of RV activity induced signals to the position and motion of the active regions on the visible stellar disc, creating complex patterns \citep[see e.g.,][]{Dumusque2014}.
Fortunately, this complexity can be described, as a first order approach, as the derivative of the function $G(t)$ that describes the area covered by active regions on the stellar surface as a function of time \citep[][]{Aigrain2012,Rajpaul2015}.
This can be seen empirically in this system as the recovered amplitudes for the GP describing the RV time-series are $A_{\rm RV}$=\jAzero\,\ms\ (with a posterior truncated at zero, see Fig~\ref{fig:correlations}) and $B_{\rm RV}$=\jAone\,\ms\,d, i.e a significant detection (see Fig~\ref{fig:correlations}).  
{We can also see that the process describing the stellar signal in the \sshk\ time-series has an amplitude of \jAtwo\ that is significantly different from zero. This suggest that the stellar signal in the \sshk\ is also constrained with our model (for more discussion on this see Appendix~\ref{sec:shk}).}
This demonstrates that for \hd, the RV signal is mainly described by $\dot{G}(t)$, while \sshk\ time-series is well described by $G(t)$. We note that this behaviour has been observed empirically in other young stars that show high harmonic complexity \citep[e.g,][]{Barragan2019}.
For a further discussion on the importance of the derivative of the GPs when dealing with high harmonic complexity see \citet{pyaneti2}.
{In Appendix~\ref{sec:gptests} we show further tests that we perform in order to ensure that our stellar modelling is robust.}

As we mention in Sect.~\ref{sec:multigp}, we assume there is no chromatic variation of the stellar signal between our four different spectrographs. 
From Figure~\ref{fig:timeseries} we can see that PFS and HIRES data overlap with some HARPS observations. Figure~\ref{fig:timeseries} shows that the PFS and HIRES observations are consistent with the same time-scales and amplitudes as the HARPS data set. This encourages that our assumption of describing different instruments that observe in similar wavelengths with the same underlying function is correct.
We note that the CORALIE observations were taken months before the rest of our RV data, and given the \lbe $ = $ \jlambdae\,d, we expect that \hd\ had a different configuration of active regions at that time. However, we note that the CORALIE observations seem to have the same amplitudes and scales as the RVs of the other instruments.






\subsection{Dynamical analysis}

We performed an orbital stability analysis of the \hd\ system using the software \texttt{mercury6} \citep{Chambers1999}. We assume that both planets have co-planar orbits and we use our derived parameters in Table~\ref{tab:derived} to create our \texttt{mercury6} set-up. 
We evolved the system for 1 Gyr with steps of 0.5\,d per integration. For \hdb\ we found negligible changes on the orbital parameters of the planet, except for the eccentricity that fluctuated with a maximum change of 0.08. 
For \hdc\ we found changes of its eccentricity $< 0.08$ and a maximum variation of $5\times10^{-5}$\,AU in its semi-major axis.
Therefore, we conclude that the orbital and planetary parameters derived for \hd\ are consistent with a dynamically stable system.

\subsection{Exoplanet compositions}

\begin{figure*}
    \centering
    \includegraphics[width=0.9\textwidth]{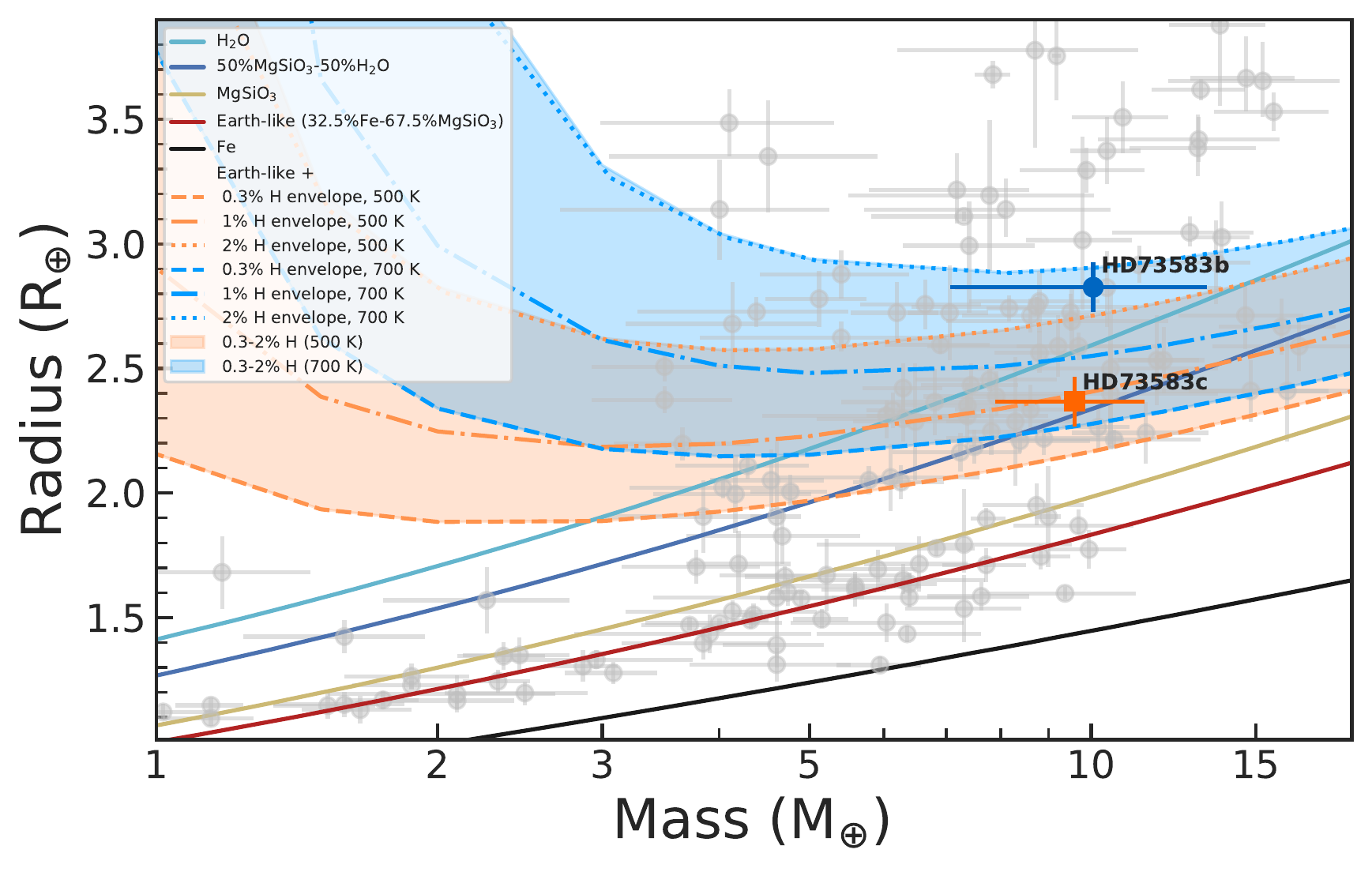}
    \caption{Mass \emph{vs} radius diagram for small exoplanets ($1 < R_{\rm p} < 4\, R_\oplus$ and $1 < M_{\rm p} < 20\, M_\oplus$). Grey points with error bars show planets with mass and radius measurements better than 30\% \citep[As in the TEPCAT catalogue, \url{https://www.astro.keele.ac.uk/jkt/tepcat/}, ][]{Sotuhworth2007}. \hdbc\ are shown with a (blue) circle and a (orange) square, respectively. Solid lines represent two-layer models as given by \citet{Zeng2016} with a different colour corresponding to a different mixture of elements. Non-solid lines correspond to rocky cores surrounded by an Hydrogen envelope with 0.3\% (dashed line), 1\% (dash-dotted line), and 2\% (dotted line) Hydrogen mass for exoplanets with equilibrium temperatures of 500\,K (orange, similar to \hdc's  $T_{\rm eq}$) and 700\,K (blue, similar \hdb's $T_{\rm eq}$) as given by \citet{Zeng2019}.}
    \label{fig:mr}
\end{figure*}

Figure~\ref{fig:mr} shows a mass-radius diagram for small exoplanets ($1 < R_{\rm p} < 4\, R_\oplus$ and $1 < M_{\rm p} < 20 M_\oplus$) detected with a precision better than 30\% in radius and mass. 
We also over-plot the two layer exoplanet models by \citet{Zeng2016}, together with the Earth-like {interior} plus Hydrogen envelope models given by \citet{Zeng2019}.
With a mass of \mpb\ and radius of \rpb, \hdb\  has a density of \denpb. \hdb\ lies above the pure water composition model. This implies that some percentage of the planet radius has to be gaseous \citep[see e.g.,][]{Russell2021}.
With an equilibrium temperature of \Teqb\, \hdb\ is consistent with a composition made of an Earth-like {interior} with a thick Hydrogen envelope accounting for approximately 2\% of the planet's mass. Nonetheless, we note that there is a degeneracy on determining exoplanet compositions in a mass-radius diagram. For example, {it is also possible to explain \hdb\ with a water rich interior with an Hydrogen envelope that accounts for only 0.3\% of the planet's mass} \citep[we do not show such models in Fig.~\ref{fig:mr}, for more details see][]{Zeng2019}.
The other planet, \hdc, has a mass of \mpc\ similar to \hdb\ but a significantly smaller radius of \rpc. This gives a bulk density of \denpc for \hdc, that puts it below the pure water composition model. 
Therefore, \hdc\ is consistent with a solid water-rich world made of 50\% water ice and 50\% silicates. However, we can see that \hdc\ is also consistent with a planet made of a Earth-like {interior}, surrounded by a Hydrogen envelope that could account for 1\% of its mass (taking into account its equilibrium temperature of \Teqc).

\begin{figure}
    \centering
    \includegraphics[width=0.49\textwidth]{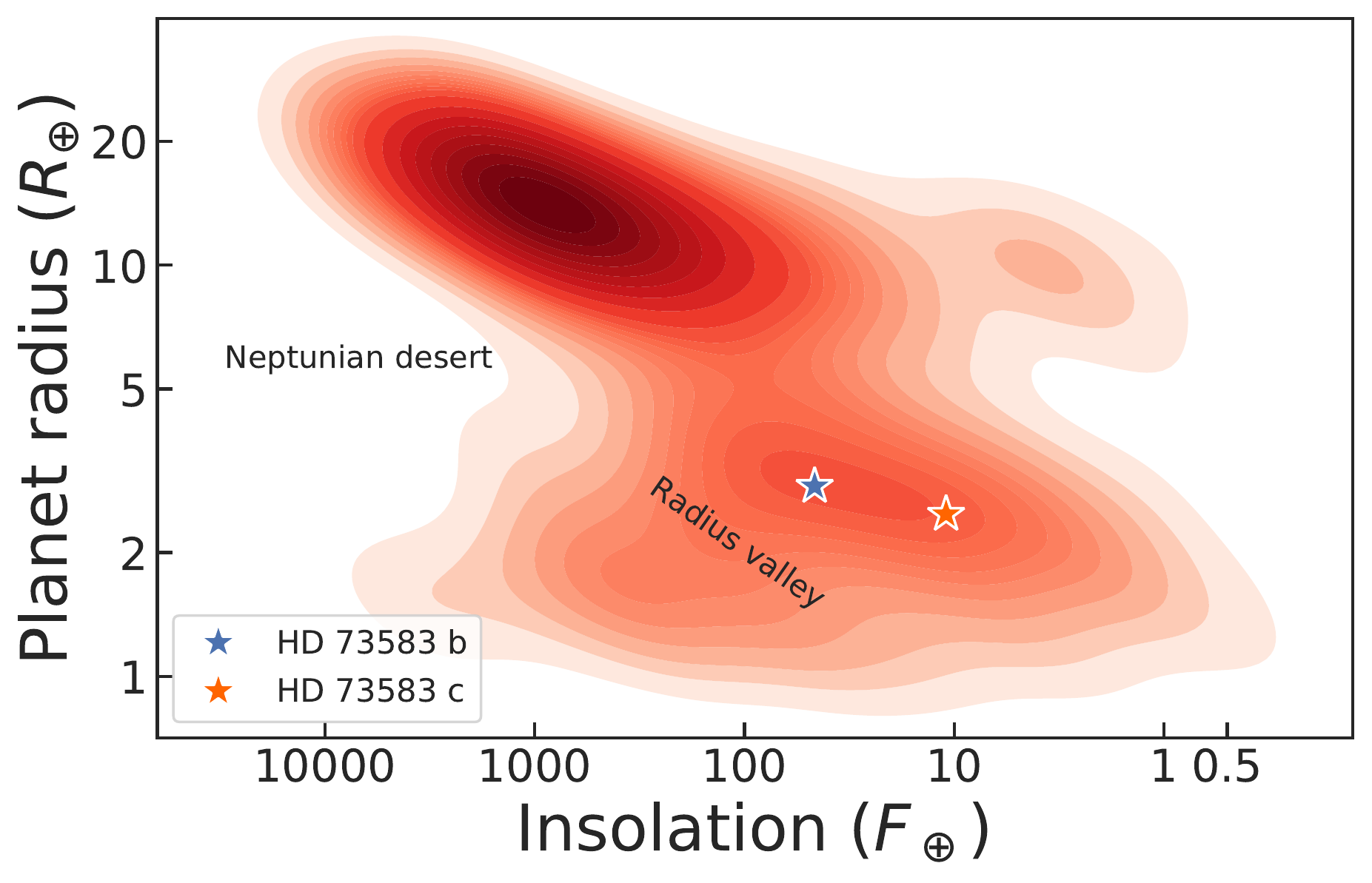}
    \caption{Insolation vs planetary radius diagram. Red contours show the occurrence of transiting exoplanets (As in the TEPCAT catalogue). \hdb\ and \hdc\ are represented with a blue and orange star, respectively. The approximately locations of the "hot Neptunian desert" and "radius valley" are shown for guidance.}
    \label{fig:insolation}
\end{figure}

Figure~\ref{fig:insolation} shows an insolation vs planetary radius diagram with the approximately positions of the Neptunian desert and radius valley indicated with text.
It is worth to note that \hdbc\ lie well above the radius valley. We therefore expect that both planets have a volatile envelope rather than being solid worlds, as suggested by previous works \citep{Fulton2017,VanEylen2018}.
Despite the degeneracy in composition for both planets,  for the remainder of the discussion in this manuscript we will assume that \hdbc\ have an Earth-like {interior} surrounded by a Hydrogen rich volatile envelope. 
A discussion of the planetary characteristics assuming different composition scenarios is out of the scope of this paper.

\begin{table}
\begin{center}
  \caption{Derived parameters for the \hd\ planets. \label{tab:derived}}
  \begin{tabular}{lcc}
  \hline
  \hline
  Parameter & \hdb's  & \hdc's  \\
   & values & values \\
  \hline
  \noalign{\smallskip}
    Planet mass $M_\mathrm{p}$ ($M_{\rm \oplus}$) &  \mpb[]  &  \mpc[]  \\
    Planet radius $R_\mathrm{p}$ ($R_{\rm \oplus}$) &  \rpb[] &  \rpc[] \\
    Planet density $\rho_{\rm p}$ (g\,cm$^{-3}$) &  \denpb[] &  \denpc[] \\
    Scaled semi-major axis  $a/R_\star$ &  \arb[]  &  \arc[]  \\
    Semi-major axis  $a$ (AU) &  \ab[] &  \ac[] \\
    Eccentricity $e$  & \eb[] & \ec[] \\
    angle of periastron $\omega_\star$ & \wb[] & \wc[] \\
    Orbit inclination $i_\mathrm{p}$ ($^{\circ}$) &  \ib[] &  \ic[] \\
    Transit duration $\tau_{14}$ (hours) & \ttotb[]  & \ttotc[]  \\
    RV at mid-transit time (\kms) & \prvb[]  & \prvc[] \\
    Planet surface gravity $g_{\rm p}$ (${\rm cm\,s^{-2}}$)$^{(a)}$ & \grapb[] & \grapc[] \\
    Planet surface gravity $g_{\rm p}$ (${\rm cm\,s^{-2}}$)$^{(b)}$ & \grapparsb[] & \grapparsc[] \\
    Equilibrium temperature  $T_\mathrm{eq}$ (K)$^{(c)}$  &   \Teqb[] &   \Teqc[] \\
    Received irradiance ($F_\oplus$) & \insolationb[] & \insolationc[] \\
    TSM$^{(d)}$ & \tsmb[] & \tsmc[] \\
   \noalign{\smallskip}
  \hline
  \multicolumn{3}{l}{$^a$ Derived using $g_{\rm p} = G M_{\rm p} R_{\rm p}^{-2}$.}\\
  \multicolumn{3}{l}{$^b$ Derived using sampled parameters following \citet{Sotuhworth2007}.}\\
  \multicolumn{3}{l}{$^c$ Assuming a zero albedo.}\\
  \multicolumn{3}{l}{$^d$ Transmission spectroscopy metric (TSM) by \citet{Kempton2018}.}\\
  \end{tabular}
\end{center}
\end{table}

We note that \hdb\ has a lower density than \hdc, but they both have similar masses.
According to \citet{Zeng2019}, planets with the same mass and same volatile Hydrogen content would have different radii (hence different density) if their temperatures are significantly different{, being} a hotter planet more bloated due to thermal inflation. We note that \hdb\ is $\sim 200$\,K hotter than \hdc\ and we would then expect a larger radius for it. However, the difference in temperature between \hdbc\ is not enough to explain the $\sim 0.5 R_\oplus$ difference in radii between both planets.
The difference in radii can be explained by extra Hydrogen content in the atmosphere of \hdb, with respect to \hdc\  (see. Fig~\ref{fig:mr}).
This is unexpected if we assume that this system has been shaped by photo-evaporation, in which the innermost planet is expected to have a more depleted atmosphere  \citep[e.g.,][]{Lopez2014,Owen2013}. In Fig.~\ref{fig:insolation} we can see that both planets lie far from the high irradiated hot Neptunian desert and the radius valley regions. This suggests that the photo-evaporation process may be slow. In this low irradiation regime, core-powered mass loss mechanisms could also play an important role sculpting the planet's atmospheres \citep{Gupta2021}. Given their youth, both planets could still evolving and experiencing atmospheric mass loss. 

\subsection{Atmospheric characterisation perspectives}

\subsubsection{Transmission spectroscopy generalities}

{We emphasise that from now on our analyses and conclusions assume that \hdbc\ are Earth-like interior + H/He envelope planets.}
Because of their youth, extended atmosphere, and host star brightness, \hdbc\ are excellent candidates to perform transmission spectroscopy. 
We note that \hdb\ has a Transmission spectroscopic metric (TSM) of \tsmb[], that is well above the threshold at 90 suggested by \citet{Kempton2018}. Therefore, this target is highly valuable target for the James Webb Space Telescope (JWST).
We note that \hdc's TSM value of \tsmc\ is below the threshold. However, we discuss here the atmospheric study perspectives for both planets.

\begin{figure}
    \centering
    \includegraphics[width=0.48\textwidth]{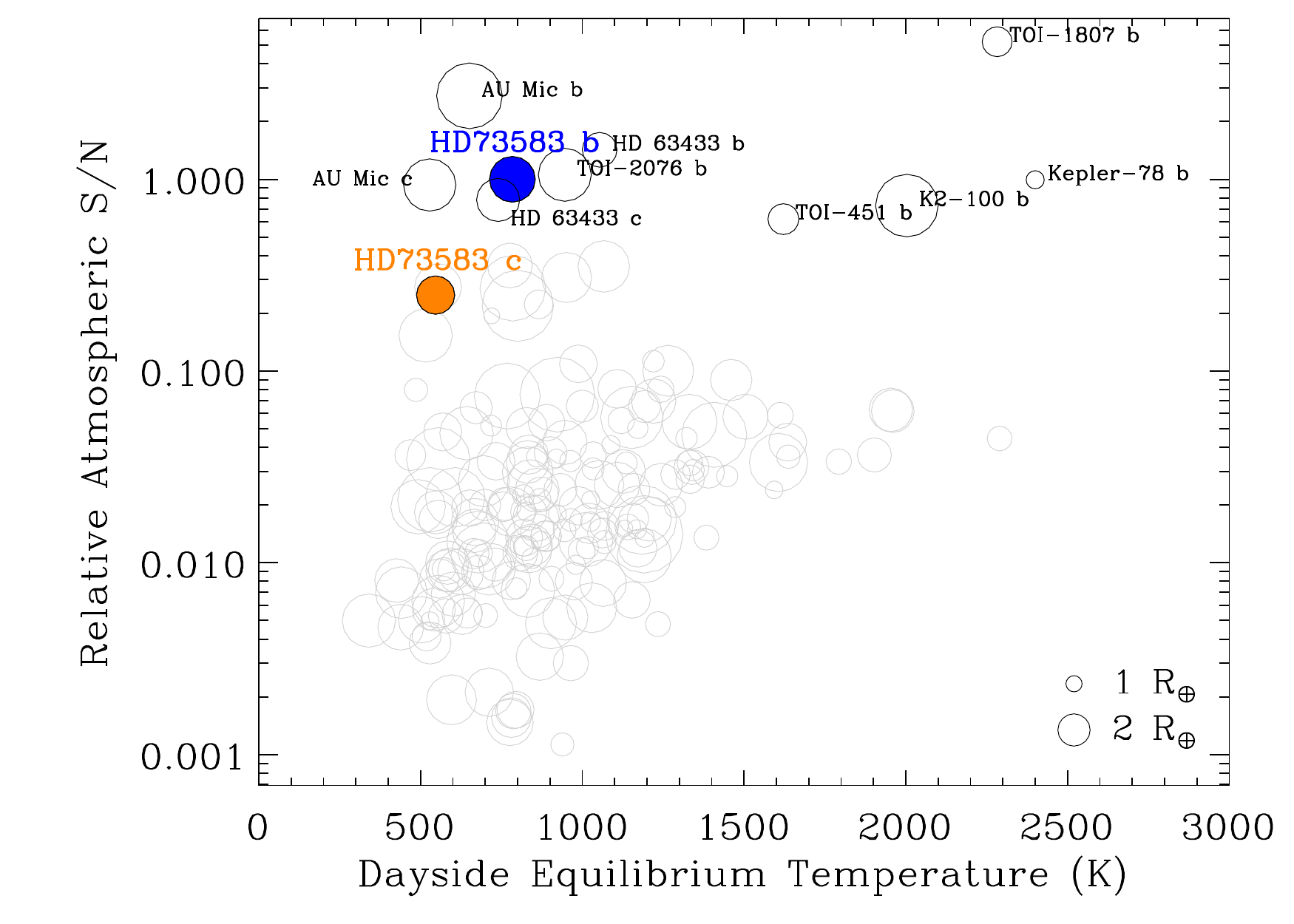}
    \caption{Relative S/N of an atmospheric signal for all young ($< 1$\,Gyr) exoplanet candidates with $R_{\rm p} < 5 R_\oplus$. Circles size show the planetary radius.
    \hdb\ and \hdc\ are shown with a filled blue and orange circles, respectively. \hdb\ is used as the S/N reference.}
    \label{fig:snratmos}
\end{figure}

Figure~\ref{fig:snratmos} displays a relative atmospheric detection S/N metric (normalised to \hdb) for all well-characterised young transiting planets with $R < 5\,R_\oplus$. 
The sample is taken from the NASA Exoplanet Archive. 
The atmospheric signal is calculated in a similar way in \citet{Niraula2017}. The atmospheric signal is dominated by the atmospheric scale height, favouring hot, extended atmospheres, and the host star radius, favouring small, cool stars. 
The relative S/N calculation scales with properties that make it favourable to detect and measure this signal. Our metric is similar to the TSM in \citet{Kempton2018}. The difference with our metric, is that instead of calculating this per transit, we calculate it based on time, thus adding a $P^{-0.5}$ term. Given the observational challenges of observing planets in transit with highly oversubscribed facilities, the frequency of transits is a very important constraint on obtaining atmospheric measurements. We assume an effective scale height \citep[$h_{\rm eff} = 7$ Hill radii;][]{MillerRicci2009} using the equilibrium temperature, a Bond albedo of $\alpha=0.3$, and an atmospheric mean molecular weight of $\mu = 20$. Because this is a relative assessment, and we are assuming identical properties for all the atmospheres in this sample, the precise value of these variables do not change the results. 
Table~\ref{tab:snratmos} shows the top ten best young planets candidates in terms of expected S/N. \hdb\ occupies the fifth position in this rank, making it an excellent target for transmission spectroscopy follow-up efforts. Even if not showed in Table~\ref{tab:snratmos}, We note that \hdc\ lies in the sixteenth position in the same ranking.
It is worth to mention that the scientific value of \hdbc\ is even higher given that both planets form part of the same system. This will allow to perform comparative atmospheric composition and mass loss studies that are crucial to test theoretical models.

\begin{table}
    \centering
    \caption{Top ten best young exoplanets candidates for transmission spectroscopy S/N normalised to \hdb, that is highlighted in bold.}
    \begin{tabular}{llccccc}
    \hline
    \hline
Rank     &   Name   &       $T_{\rm eq}$  &  S/N relative to  & $R_\star$ & Orbital \\
     &     &       atmosphere &  \hdb\ &  &  Period \\
\hline
 1        &    TOI-1807 b & 2282.4 & 5.217 & 0.68 &  0.5494 \\
 2        &      AU Mic b &  651.2 & 2.725 & 0.75 &  8.4630 \\
 3        &    HD 63433 b & 1053.7 & 1.425 & 0.91 &  7.1079 \\
 4        &    TOI-2076 b &  945.9 & 1.056 & 0.76 & 10.3557 \\
 5        &     {\bf HD 73583 b} & {\bf 783.9} & {\bf 1.000} & {\bf 0.66} &  \bf{6.3980} \\
 6        &   Kepler-78 b & 2400.0 & 0.994 & 0.75 &  0.3550 \\
 7        &      AU Mic c &  528.5 & 0.934 & 0.75 & 18.8590 \\
 8        &    HD 63433 c &  739.8 & 0.781 & 0.91 & 20.5453 \\
 9        &      K2-100 b & 2003.3 & 0.729 & 1.24 &  1.6739 \\
10        &     TOI-451 b & 1621.6 & 0.620 & 0.88 &  1.8587 \\
\hline
    \end{tabular}
    \label{tab:snratmos}
\end{table}

\subsubsection{Hydrogen escape}
\begin{figure}
    \centering
    \includegraphics{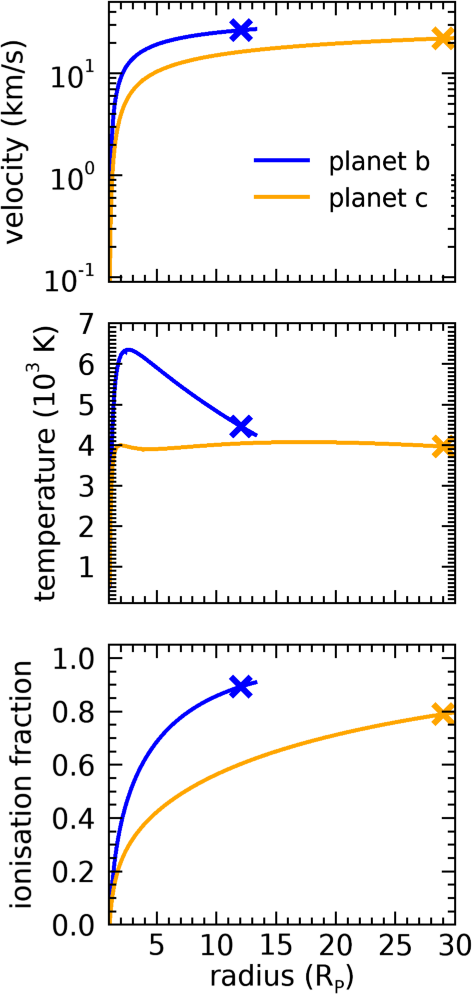}
    \caption{Atmospheric profiles derived from our hydrodynamic escape simulations showing the radial (i.e., outwards) velocity of the planetary outflow (top), its temperature (middle) and ionisation fraction (bottom). Crosses indicate the radial distance to the Roche lobe. \hdb\ and \hdc\ properties are shown with blue and orange lines, respectively.}
    \label{fig:hescape}
\end{figure}

Using the 1D hydrodynamic escape model described in \citet{Allan2019}, we predict the planetary upper atmosphere properties, such as the evaporation rate of the two planets and their velocity and density atmospheric structure.
We consider an atmosphere that is made only of Hydrogen, and the ionisation balance is self-consistently derived by including  Ly-alpha cooling and photoionisation by XUV stellar irradiation \citep{Allan2019}. To derive the XUV stellar flux, we proceed as follows. 
From the median value of HARPS $\rm log\,R^{\prime}_{\mathrm{HK}}$ (-4.465 $\pm$ 0.015), we calculated the Ca\,II\,H\&K chromospheric emission flux using the equations in \citet{fossati2017} and derived the XUV flux using the scaling relations of \citet{linsky2013} and \citet{linsky2014}. The corresponding stellar XUV luminosity is  $9 \times 10^{-6}~L_\odot$, which results in fluxes of $3.5 \times 10^3$ and $8.3 \times 10^2$ ${\rm erg\,cm^{-2}\,s^{-1}}$ at the orbital distance of planets b and c, respectively. Without knowledge of the spectral energy distribution in the X-ray and EUV bands, we assume this flux is concentrated at 20\,eV as done in \citet{Allan2019} \citep*[see also][]{Hazra2020}.
The radial velocity of the escaping atmosphere, temperature and ionisation fraction for both planets are shown in Fig.~\ref{fig:hescape}. 
The simulations are computed up to the Roche lobe (indicated by the crosses in Figure~\ref{fig:hescape}). At these points, we see that the escaping atmospheres reach velocities between $\sim 20$ and $\sim 30$ \kms. These atmospheres are 50\% ionised at distances of $\sim$ 2.8 $R_{\rm p}$ and $\sim$ 7.4 $R_{\rm p}$, for planets b and c, respectively. The 50\% threshold is achieved further out for planet c, because of the lower stellar flux it receives. We found evaporation rates for \hdb\ of $2.4 \times 10^{10}\,{\rm g\,s^{-1}}$ and for \hdc\ of $5.4 \times 10^{9}\,{\rm g\,s^{-1}}$.

\begin{figure}
    \centering
    \includegraphics[width=0.4\textwidth]{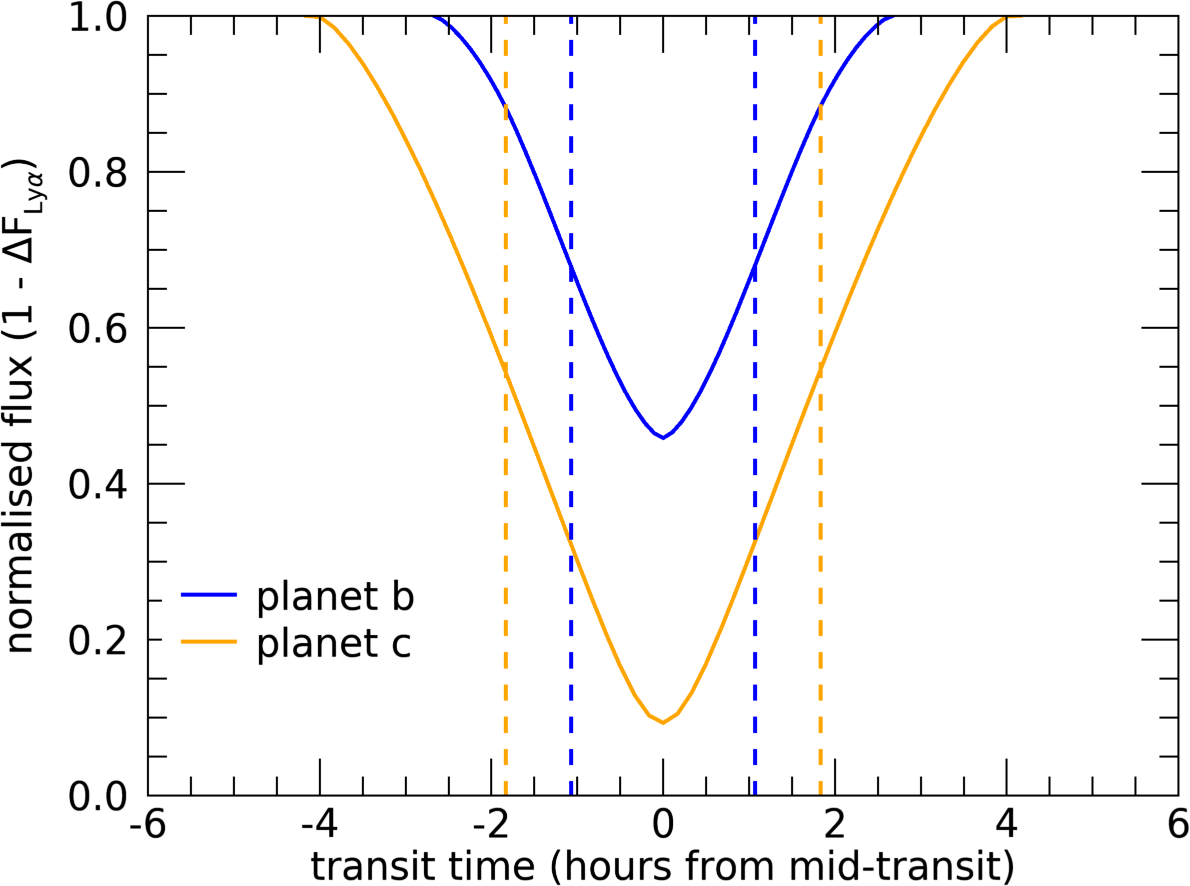}\\
    \includegraphics[width=0.4\textwidth]{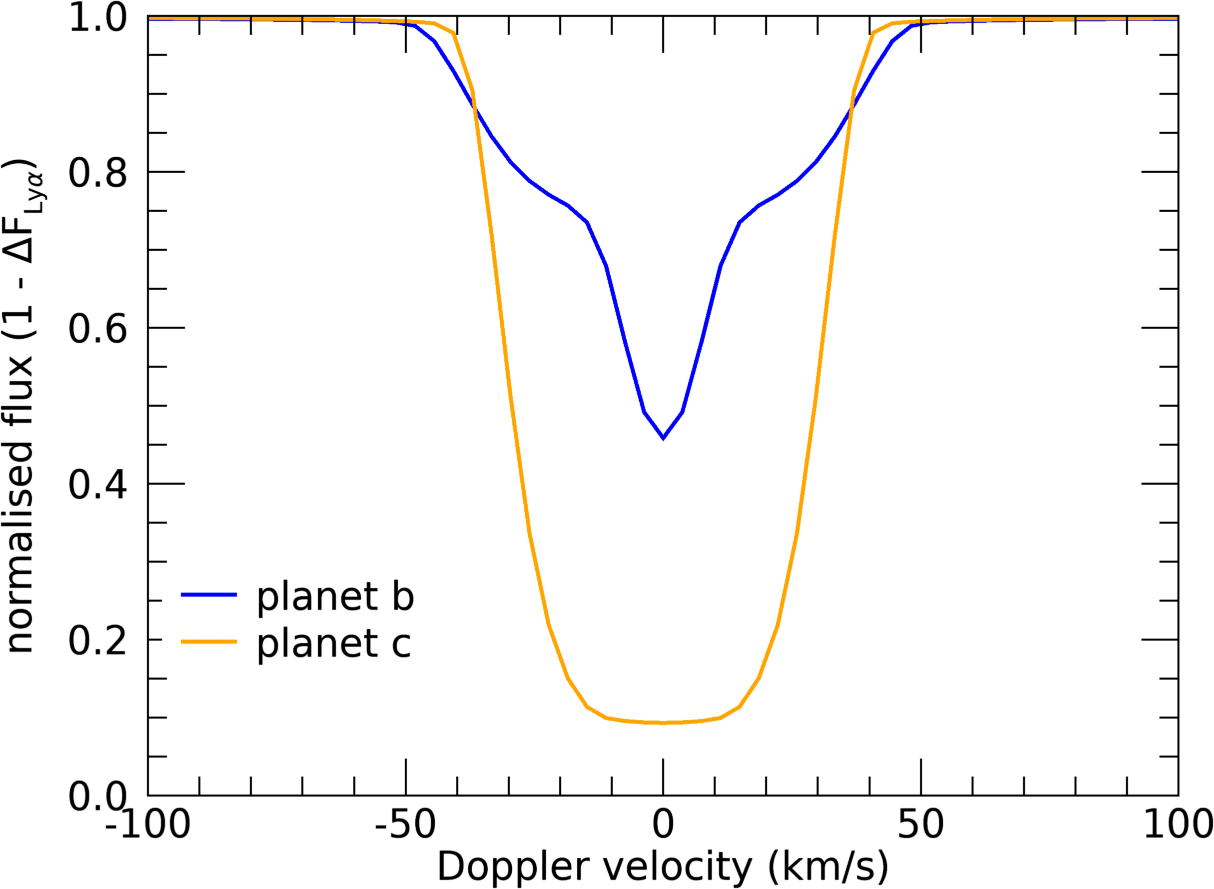}
    \caption{Top: Ly$\alpha$ light curves for planet \hdb\ (blue) and \hdc\ (orange). Dashed lines mark the duration of the geometric transit. Bottom: Normalised absorption profile at mid-transit for planet \hdb\ (blue) and \hdc\ (orange). }
    \label{fig:lya_lightcurve}
\end{figure}

From the neutral Hydrogen density, velocity and temperature profiles, we can predict the transit in Ly$\alpha$ and H$\alpha$ for both planets. This is done using a ray tracing model \citep{2018MNRAS.481.5296V, Allan2019}, in which we shoot stellar rays through the planetary atmosphere and we calculate how much of these rays are transmitted through the atmosphere. The top panel of Fig. \ref{fig:lya_lightcurve} shows the predicted light curves at the Ly$\alpha$ line centre. We found a total absorption at mid-transit of 96\% for planet \hdb\ and 52\% for planet \hdc\ in Ly$\alpha$.
Although this is a strong absorption, we note from the bottom panel that this absorption is mostly concentrated in the line centre (ie, not extending too much to the line wings), {where observations are not possible due to ISM absorption and geocoronal contamination. Therefore, these estimations should be taken carefully}. 
The reason for {the large absorption at line center} is that our 1D hydrodynamic model cannot include 3D effects that could broaden the absorption.
A broader absorption could be possible if we were to include the effect of the stellar wind in our escape models \citep{Villarreal2018, Carolan2021} and other process like charge exchange and radiation pressure \citep{Bourrier2013, Khodachenko2017, Esquivel2019}, which are out of the scope of this paper.

While most of the neutral Hydrogen is found in the ground state, a fraction of the atoms are in the first excited state ($n=2$). These atoms can then absorb stellar H$\alpha$ photons, which could generate a detectable H$\alpha$ transit \citep{Jensen2012,Jensen2018, Casasayas-Barris2018, Cabot2020, Chen2020,Yan2021}.

To compute the level 2 population in the atmosphere of \hdbc, we follow the method described in \citet{Villarreal2021}. We take as an input the electron density and the temperature of the planetary atmosphere from the 1D hydrodynamic model and include the stellar Ly$\alpha$ flux as an external radiation field. This flux is approximated with a black-body function at a temperature of 8000K  \citep[see also][]{Christie2013, Huang2017}. We then use our ray tracing method and predict a small percentage of absorption during transit in H$\alpha$, 0.32\% and  0.26 \% for planets \hdbc, respectively.
This low value of absorption is consistent with the fact that, so far, hot-Jupiter like exoplanets with H$\alpha$ absorption detection  have T$_{\rm eq}>1000$\,K \citep{Jensen2012,Jensen2018,Chen2020}  while both planets studied here show relatively small equilibrium temperatures. 

\subsubsection{Helium escape}

We use a different 1D hydrodynamic escape model to estimate the absorption signatures that could be expected from HD~73583~b and HD~73583~c in the near-infrared line triplet of neutral helium at 1083~nm. The model assumes an atmosphere composed entirely of atomic Hydrogen and helium, in 9:1 number ratio. The density and velocity structures of the escaping atmosphere are based on the isothermal Parker wind model described in \citet{OklopcicHirata2018}. For the input stellar spectrum used in our model, we construct a spectral energy distribution appropriate for HD 73583 (a K4-type star) by taking an average between the spectra of $\epsilon$ Eridani (K2 type) and HD 85512 (K6 type) obtained by the MUSCLES survey \citep{France2016}. We further adjust the high-energy part of the spectrum to match the expected XUV flux of HD 73583 at the orbital distances of the two planets. In addition to atmospheric composition, the main free parameters of the model are the temperature of the escaping atmosphere and the total mass loss rate. We run a grid of models spanning a range of temperatures (3000 - 7000 K) and mass loss rates (between $\dot{\mathrm{M}} = 10^8$\,\gs\ and $\dot{\mathrm{M}} = 10^{10.5}$\,\gs). To calculate the abundance of helium atoms in the excited (metastable) state which is responsible for the 1083\,nm absorption line, we perform radiative transfer calculations for a 1D atmospheric profile along the planet's terminator. We compute the transmission spectrum for a planet at mid-transit, assuming the planet is tidally locked and the atmosphere is rotating with the planet as a solid body.  

\begin{figure}
    \centering
    \includegraphics[width=0.5\textwidth]{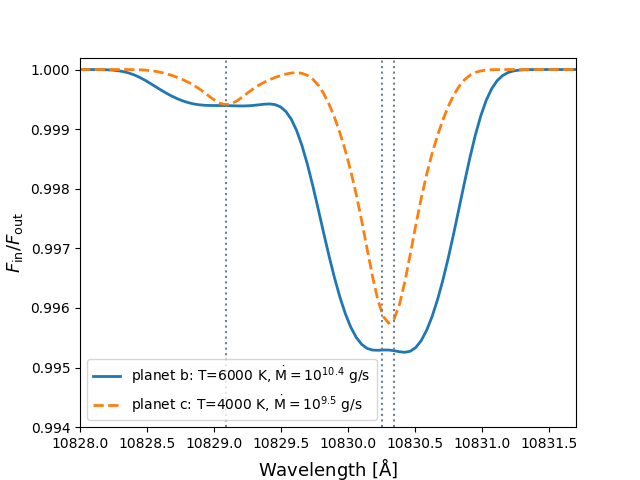}
    \caption{Calculated excess absorption at mid-transit in the helium 1083~nm line for HD 73583 b (solid blue line) and HD 73583 c (dashed orange line). The values of the input parameters are motivated by the results of the atmospheric escape simulations using pure Hydrogen composition. Dotted lines mark the wavelengths (in air) of the helium triplet lines.}
    \label{fig:helium}
\end{figure}

The predicted signals for both planets are typically small, below 1\% excess absorption at the centre of the 1083 nm line. Figure \ref{fig:helium} shows the expected excess absorption at midtransit for the mass loss rates and temperatures similar to those predicted by the previously described Hydrogen simulations. We note that the 1D models used for simulating Hydrogen and helium signals use different assumptions and atmospheric profiles. Using the results of one model as input for the other is not entirely self-consistent, so these results should only be considered as rough estimates of the helium absorption signals from these two planets.

{We note that while this paper  was under review, \citet[][]{Zhang2021} reported Helium detection in the atmosphere of \hdb. This confirms the expected atmospheric evolution on this system and encourages further spectroscopic follow-up.
A comparison of our results with their detection and models is out of the scope of this manuscript.
}

\section*{Acknowledgements}
{\footnotesize
This work was supported by the KESPRINT collaboration, an international
consortium devoted to the characterisation and research of exoplanets
discovered with space-based missions (www.kesprint.science).
{We thank the referee for their helpful comments and suggestions that improved the quality of this manuscript.}
We acknowledge the use of public TESS data from pipelines at the TESS Science Office and at the TESS Science Processing Operations Center. Resources supporting this work were provided by the NASA High-End Computing (HEC) Program through the NASA Advanced Supercomputing (NAS) Division at Ames Research Center for the production of the SPOC data products.
This work makes use of observations from the LCOGT network. Part of the LCOGT telescope time was granted by NOIRLab through the Mid-Scale Innovations Program (MSIP). MSIP is funded by NSF.
This paper is in part based on data collected under the NGTS project at the ESO La Silla Paranal Observatory.  The NGTS facility is operated by the consortium institutes with support from the UK Science and Technology Facilities Council (STFC)  projects ST/M001962/1 and  ST/S002642/1.
This research has made use of the NASA Exoplanet Archive, which is operated by the California Institute of Technology, under contract with the National Aeronautics and Space Administration under the Exoplanet Exploration Program.
Some of the observations in the paper made use of the High-Resolution Imaging instrument Zorro obtained under Gemini LLP Proposal Number: GN/S-2021A-LP-105. Zorro was funded by the NASA Exoplanet Exploration Program and built at the NASA Ames Research Center by Steve B. Howell, Nic Scott, Elliott P. Horch, and Emmett Quigley. Zorro was mounted on the Gemini North (and/or South) telescope of the international Gemini Observatory, a program of NSF’s OIR Lab, which is managed by the Association of Universities for Research in Astronomy (AURA) under a cooperative agreement with the National Science Foundation. on behalf of the Gemini partnership: the National Science Foundation (United States), National Research Council (Canada), Agencia Nacional de Investigaci\'on y Desarrollo (Chile), Ministerio de Ciencia, Tecnolog\'ia e Innovaci\'on (Argentina), Minist\'erio da Ci\^encia, Tecnologia, Inovações e Comunicações (Brazil), and Korea Astronomy and Space Science Institute (Republic of Korea).
OB, BK, and SA acknowledge that this publication is part of a project that has received funding from the European Research Council (ERC) under the European Union’s Horizon 2020 research and innovation programme (Grant agreement No. 865624).
DG and LMS gratefully acknowledge financial support from the \emph{Cassa di Risparmio di Torino} foundation under Grant No. 2018.2323 ``Gaseous or rocky? Unveiling the nature of small worlds''.
D.J.A acknowledges support from the STFC via an Ernest Rutherford Fellowship (ST/R00384X/1).
APH and ME acknowledges grant HA 3279/12-1  within the DFG Schwerpunkt SPP 1992,
``Exploring the Diversity of Extrasolar Planets''.
JS and PK would like to acknowledge support from MSMT grant LTT-20015.
We acknowledges the support by FCT - Fundação para a Ciência
e a Tecnologia through national funds and by FEDER through COMPETE2020
- Programa Operacional Competitividade e Internacionalização
by these grants: UID/FIS/04434/2019; UIDB/04434/2020; UIDP/04434/2020;
PTDC/FIS-AST/32113/2017 \& POCI-01-0145-FEDER-032113; PTDC/FISAST
/28953/2017 \& POCI-01-0145-FEDER-028953.
A.De. acknowledges the financial support of the European Research Council (ERC) under the European Union's Horizon 2020 research and innovation programme (project {\sc Four Aces}; grant agreement No 724427). A.De. also acknowledges financial support of the the Swiss National Science Foundation (SNSF) through the National Centre for Competence in Research ``PlanetS''.
MF, IYG, JK and CMP gratefully acknowledge the support of the  Swedish National Space Agency (DNR 177/19, 174/18, 2020-00104, 65/19).
FGC thanks the Mexican national council for science and technology (CONACYT, CVU-1005374).
MS acknowledge the financial support of the Inter-transfer grant no LTT-20015.
J.L-B. acknowledges financial support received from ”la Caixa” Foundation (ID 100010434) and from the European Union’s Horizon 2020 research and innovation programme under the Marie Skłodowska-Curie grant agreement No 847648, with fellowship code LCF/BQ/PI20/11760023.
AAV acknowledges funding from the European Research Council (ERC) under the European Union's Horizon 2020 research and innovation programme (grant agreement No 817540, ASTROFLOW).
J.M.A.M. is supported by the National Science Foundation Graduate Research Fellowship Program under Grant No. DGE-1842400. J.M.A.M. acknowledges the LSSTC Data Science Fellowship Program, which is funded by LSSTC, NSF Cybertraining Grant No. 1829740, the Brinson Foundation, and the Moore Foundation; his participation in the program has benefited this work.
R.A.R. is supported by the NSF Graduate Research Fellowship, grant No. DGE 1745301.
R.L. acknowledges financial support from the Spanish Ministerio de Ciencia e Innovación, through project PID2019-109522GB-C52, and the Centre of Excellence "Severo Ochoa" award to the Instituto de Astrofísica de Andalucía (SEV-2017-0709).
PC acknowledges the generous support from Deutsche Forschungsgemeinschaft
(DFG) of the grant CH 2636/1-1.
SH acknowledges CNES funding through the grant 837319.
V.A. acknowledges the support from Funda\c{c}\~ao para a Ci\^encia e Tecnologia (FCT) through Investigador FCT contract nr. IF/00650/2015/CP1273/CT0001.
O.D.S.D. is supported in the form of work contract (DL 57/2016/CP1364/CT0004) funded by national funds through Fundação para a Ciência e Tecnologia (FCT).
A.Os. is supported by an STFC studentship.
X.D. would like to acknowledge the funding from the European Research Council (ERC) under the European Union’s Horizon 2020 research and innovation programme (grant agreement SCORE No 851555).
H.J.D. acknowledges support from the Spanish Research Agency of the Ministry of Science and Innovation (AEI-MICINN) under the grant `Contribution of the IAC to the PLATO Space Mission' with reference PID2019-107061GB-C66, DOI: 10.13039/501100011033.
D. D. acknowledges support from the TESS Guest Investigator Program grant 80NSSC19K1727 and NASA Exoplanet Research Program grant 18-2XRP18\_2-0136.
A.Ok. gratefully acknowledges support from the Dutch Research Council NWO Veni grant.}

\section*{Data Availability}

Some of the codes used in this manuscript are available as online supplementary material accessible following the links provided in the online version. Our spectroscopic measurements are available as a suplementary material in the online version of this manuscript.

\section*{Affiliations}
{\footnotesize
$^{1}$ Sub-department of Astrophysics, Department of Physics, University of Oxford, Oxford, OX1 3RH, UK  \label{oxford} \\
$^{2}$ Department of Physics, University of Warwick, Coventry CV4 7AL, UK \label{pwarwick} \\
$^{3}$ Centre for Exoplanets and Habitability, University of Warwick, Coventry CV4 7AL, UK \label{hwarwick} \\
$^{4}$ Dipartimento di Fisica, Universit\`a degli Studi di Torino, Via Pietro Giuria 1, 10125 Torino, Italy \label{torino} \\
$^{5}$ Department of Astronomy, Indiana University, Bloomington, IN 47405 \label{indiana} \\
$^{6}$ Leiden Observatory, Leiden University, PO Box 9513, 2300 RA Leiden, The Netherlands \label{leiden} \\
$^{7}$ Instituto de Astronom\'ia Te\'orica y Experimental (CONICET - UNC). Laprida 854, X500BGR. C\'ordoba, Argentina \label{cordoba} \\
$^{8}$ Anton Pannekoek Institute for Astronomy, University of Amsterdam, Science Park 904, NL-1098 XH Amsterdam, Netherlands \label{apia} \\
$^{9}$ 501 Campbell Hall, University of California at Berckeley, Berkeley, CA 94720, USA \label{berkeley} \\
$^{10}$ Department of Physics and Astronomy, University of New Mexico, 1919 Lomas Blvd NE, Albuquerque, NM 87131, USA \label{newmexico} \\
$^{11}$ Center for Astrophysics \textbar \ Harvard \& Smithsonian, 60 Garden Street, Cambridge, MA 02138, USA \label{hardvard} \\
$^{12}$ Department of Earth and Space Sciences, Chalmers University of Technology, Onsala Space Observatory, 439 92 Onsala, Sweden\label{onsala} \\
$^{13}$ Instituto de Astrof\'isica e Ci\^encias do Espa\c{c}o, Universidade do Porto, CAUP, Rua das Estrelas, 4150-762 Porto, Portugal \label{iaporto} \\
$^{14}$ Astrophysics Group, Keele University, Staffordshire ST5 5BG, UK \label{keele} \\
$^{15}$ NASA Ames Research Center, Moffett Field, CA 94035, USA \label{nasaames}  \\
$^{16}$ Astronomy Department and Van Vleck Observatory, Wesleyan University, Middletown, CT 06459, USA \label{wesleyan} \\
$^{17}$ Department of Space, Earth and Environment, Astronomy and Plasma Physics, Chalmers University of Technology, 412 96 Gothenburg, Sweden \label{chalmers} \\
$^{18}$ Geneva Observatory, University of Geneva, Chemin Pegasi 51, 1290 Versoix, Switzerland \label{geneve} \\
$^{19}$ Department of Astronomy, California Institute of Technology, Pasadena, CA 91125, USA \label{caltech} \\
$^{20}$ Carnegie Earth \& Planets Laboratory, 5241 Broad Branch Road, NW, Washington, DC 20015, USA \label{carnegie} \\
$^{21}$ SETI Institute, Mountain View, CA  94043, USA \label{seti} \\
$^{22}$ Thüringer Landessternwarte Tautenburg, Sternwarte 5, D-07778 Tautenberg, Germany \label{tautenburg} \\
$^{23}$ Mullard Space Science Laboratory, University College London, Holmbury St Mary, Dorking, Surrey RH5 6NT, UK \label{mullarducl} \\
$^{24}$ European Southern Observatory, Alonso de Cordova 3107, Vitacura, Santiago de Chile, Chile \label{lasilla} \\
$^{25}$ Departamento de F\'isica e Astronomia, Faculdade de Ci\^encias, Universidade do Porto, Rua do Campo Alegre, 4169-007 Porto, Portugal\label{dpporto} \\
$^{26}$ Department of Astronomy and Astrophysics, University of California, Santa Cruz, CA 95064, USA \label{santacruz} \\
$^{27}$ NSF Graduate Research Fellow \label{nsf} \\
$^{28}$ Centro de Astrobiología (INTA-CSIC), Camino Bajo del Castillo s/n, 28692, Villanueva de la Cañada, Madrid (SPAIN) \label{madrid} \\
$^{29}$ Department of Physics, and Institute for Research on Exoplanets, Universit\'e de Montr\'eal, Montreal, H3T 1J4, Canada \label{montreal} \\
$^{30}$ Jet Propulsion Laboratory, California Institute of Technology, 4800 Oak Grove Drive, Pasadena, CA 91109, USA \label{jpl} \\
$^{31}$ Institute of Planetary Research, German Aerospace Center, Rutherford-strasse 2, D-12489 Berlin, Germany \label{berlin} \\
$^{32}$ School of Physics and Astronomy, University of Leicester, University Road, Leicester, LE1 7RH, UK \label{leicester} \\
$^{33}$ Banting Fellow \label{banting} \\
$^{34}$ McDonald Observatory and Center for Planetary Systems Habitability, University of Texas, Austin TX 78712 USA \label{mcdonald} \\
$^{35}$ Carnegie Observatories, 813 Santa Barbara Street, Pasadena, CA 91101, USA \label{carnegie2} \\
$^{36}$ Department of Physics and Astronomy, University of Kansas, Lawrence, KS, USA \label{kansas} \\
$^{37}$ European Space Agency (ESA), European Space Research and Technology Centre (ESTEC), Keplerlaan 1, 2201 AZ Noordwijk, The Netherlands \label{esa} \\
$^{38}$ George Mason University, 4400 University Drive, Fairfax, VA, 22030 USA \label{mason} \\
$^{39}$ Department of Physics and Kavli Institute for Astrophysics and Space Research, Massachusetts Institute of Technology, Cambridge, MA 02139, USA \label{kavlimit} \\
$^{40}$ Department of Astrophysical Sciences, Peyton Hall, 4 Ivy Lane, Princeton, NJ 08544, USA \label{princeton} \\
$^{41}$ Instituto de Astrof\'{i}sica de Canarias, 38205 La Laguna, Tenerife, Spain \label{iac} \\
$^{42}$ Departamento de Astrof\'isica, Universidad de La Laguna, E-38206 La Laguna, Spain \label{laguna} \\
$^{43}$ European Southern Observatory, Alonso de Cordova, Vitacura, Santiago, Chile \label{eso} \\
$^{44}$ NASA Exoplanet Science Institute, Caltech/IPAC, Mail Code 100-22, 1200 E. California Blvd., Pasadena, CA 91125, USA \label{nasacaltech} \\
$^{45}$ Departamento de Astronom\'ia, Universidad de Guanajuato, Callej\'on de Jalisco s/n, 36023, M\'exico \label{guanas} \\
$^{46}$ Gemini Observatory/NSF's NOIRLab, 670 N. A'ohoku Place, Hilo, HI 96720, USA \label{gemini} \\
$^{47}$ Astrobiology Center, 2-21-1 Osawa, Mitaka, Tokyo 181-8588, Japan \label{atokyo} \\
$^{48}$ National Astronomical Observatory of Japan, NINS, 2-21-1 Osawa, Mitaka, Tokyo 181-8588, Japan \label{ntokyo} \\
$^{49}$ Aix Marseille Univ, CNRS, CNES, LAM, Marseille, France \label{marseille} \\
$^{50}$ Astronomical Institute of the Czech Academy of Sciences, Fri\v{c}ova 298, 25165, Ond\v{r}ejov, Czech Republic \label{fricova} \\
$^{51}$ Centro de Astrobiolog\'ia (CAB, CSIC-INTA), Depto. de Astrof\'isica, ESAC campus, 28692, Villanueva de la Ca\~nada (Madrid), Spain \label{esac} \\
I$^{52}$ nstituto de Astrof\'isica de Andaluc\'ia (IAA-CSIC), Glorieta de la Astronom\'ia s/n, 18008 Granada, Spain \label{andalucia} \\
$^{53}$ NCCR/PlanetS, Centre for Space \& Habitability, University of Bern, Bern, Switzerland \label{bern} \\
$^{54}$ Hazelwood Observatory, Australia \label{hazelwood} \\
$^{55}$ Patashnick Voorheesville Observatory, Voorheesville, NY 12186, USA \label{voor} \\
$^{56}$ Department of Theoretical Physics and Astrophysics, Masaryk University, Kotl\'{a}rsk\'{a} 2, 61137, Brno, Czech Republic \label{mazaryk} \\
$^{57}$ Astronomical Institute of Charles University, V Hole\v{s}ovi\v{c}k\'ach 2, 180 00, Prague, Czech Republic \label{charles} \\
$^{58}$ Perth Exoplanet Survey Telescope, Perth, Western Australia \label{perth} \\
$^{59}$ Curtin Institute of Radio Astronomy, Curtin University, Bentley, Western Australia 6102 \label{curtin} \\
$^{60}$ Astrophysics Research Centre, School of Mathematics and Physics, Queen's University Belfast, BT7 1NN, Belfast, UK \label{belfast} \\
$^{61}$ Department of Astronomy, Tsinghua University, Beijing 100084, People's Republic of China \label{beijing}  \\
$^{62}$Seniors Data Scientist, SiteZeus \\
$^{63}$ Department of Physics, Engineering and Astronomy, S. F. Austin State University, 1936 North St, Nacogdoches, TX 75962, USA \label{faustin} 
}
%
\bibliographystyle{mnras} 
\bibliography{refs} 
%

\begin{appendix}

\section{Correlation plot}

\begin{figure*}
    \centering
    \includegraphics[width=0.9\textwidth]{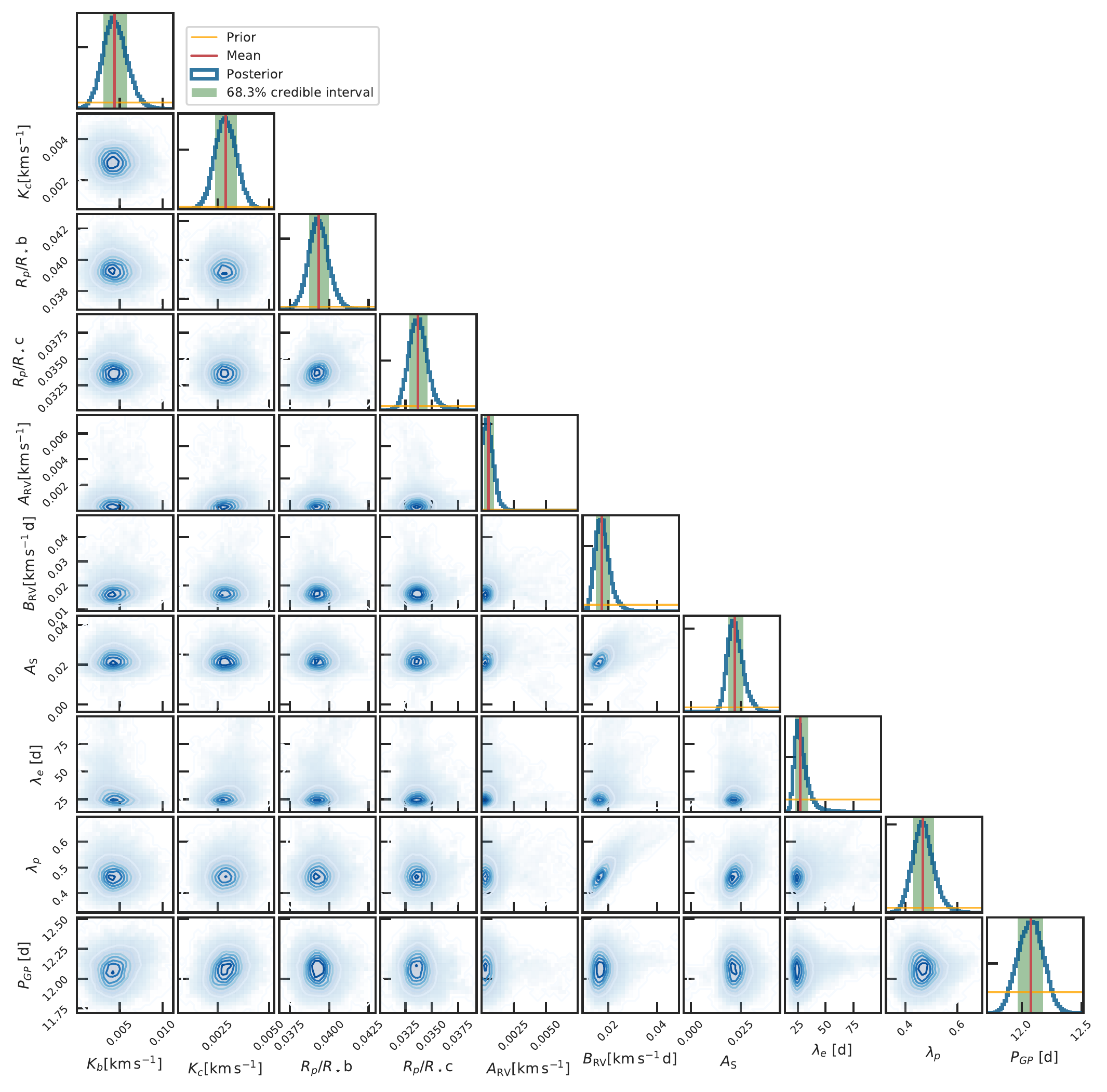}
    \caption{Posterior and correlation plots for some of the {key} sampled parameters of the joint analysis described in Sect.~\ref{sec:joint}.}
    \label{fig:correlations}
\end{figure*}

\section{Spectroscopic measurements}

\input{harpsshort}

\input{hiresshort}

\input{coralieshort}

\section{GP tests}
\label{sec:gptests}

{

In this Appendix we show some further tests that we perform in order to understand the modelling of the stellar and planetary signals in our spectroscopic time-series.

\subsection{Stellar and planetary signals in the RV time-series only}
\label{sec:rvonly}

We first make an analysis of the RV time-series to test if we can detect the planetary signals without a multi-GP approach. 
We perform a 1-dimensional GP regression using the QP kernel given in eq.~\ref{eq:gamma}.  
For the GP model we sampled for only one amplitude and the QP kernel hyper-parameters. 
The Keplerian deterministic part of our model is identical to the one described in Sec.~\ref{sec:dataanalysis}.
Our MCMC setup follows is identical to the description in Sect.~\ref{sec:joint}.  

We recover the hyper-parameters \pgp\ $=12.11_{-0.09}^{+0.08}$\,d, \lbp\ $=0.24_{-0.03}^{+0.04}$ and \lbe\ $=24_{-3}^{+4}$\,d. We note that the \pgp\ and \lbe\ are consistent with the values obtained in the joint analysis. However, \lbp\ is significantly smaller. This relative high harmonic complexity in the RV time-series is expected as we discussed at the beginning of Sect.~\ref{sec:stellarsignal} and in \citet[][]{pyaneti2}. 

The recovered planetary signals are $3.52_{-1.41}^{+1.45}$\,\ms\ and $2.63\pm0.68$\,\ms\ for \hdbc, respectively. We can see that these results are consistent with the values obtained in Sect.~\ref{sec:joint}. 
However, the results obtained with the multi-GP have better constrained values (see Table.~\ref{tab:pars}).
This improvement on the detection in the multi-GP analysis comes from a better constrain on the the underlying function $G(t)$ with help of the activity indicator \citep[For a more extensive discussion about this see][]{pyaneti2}.

\subsection{Stellar signal in the \sshk\ time-series}
\label{sec:shk}

We also perform a 1-dimensional GP regression to the \sshk\ time-series for all the instrument. This with the objective to see if we are able to recover the stellar signal purely with the activity indicator itself. For the GP model we sampled for only one amplitude and the QP kernel hyper-parameters. We also sampled for an independent offset and jitter term per each spectrograph. Our MCMC setup follows the same guidelines as the previous cases.
We recover the hyper-parameters \pgp\ $=12.22_{-0.45}^{+0.33}$\,d, \lbp\ $=0.62_{-0.14}^{+0.20}$ and \lbe\ $=31_{-09}^{+12}$\,d, and GP amplitude of $0.031_{-0.005}^{+0.007}$. These are consistent with the results obtained in Sect.~\ref{sec:dataanalysis}. This shows that the \sshk\ time-series by itself can constrain the $G(t)$ function describing the stellar signal.

It is worth to note that the recovered \pgp\ and \lbe\ in this section and in Appendix~\ref{sec:rvonly} are fully consistent.
This suggests that the stellar rotation period and spot typical lifetime manifest with the same time-scales in RV and \sshk\ observations. In contrast, the inverse of the harmonic complexity is significantly different in both cases. 
This shows how the RV and photometric-like activity indicators are not described by the same $G(t)$ signal (but they all can be described as linear combinations of a single $G(t)$ its time derivatives).

We note that the residuals of the \sshk\ time-series in Figure~\ref{fig:timeseries} present some significant variations. 
We penalise this in our model with the jitter terms. But this implies that our model may not be perfect describing our data.
This can be caused by instrumental systematics that we are not taking into account. As well as the limitation of assuming that the stellar signal can be described with a GP. However, as we show in this section, our model is enough to constrain the stellar signal at a first order.

\subsection{Tests with other activity indicators}

As a further check of our modelling of our stellar activity signal, we perform extra tests using \texttt{DRS} CCF activity indicators for our HARPS data.
The Pearson correlation coefficient between RVs with FWHM, \sshk\ and BIS are -0.16, 0.18 and -0.70, respectively. From the correlation analysis, one may think that the best activity indicator to use is the BIS span. However, it is worth to note that the stellar activity does not manifest as the same signal in the different time-series \citep[see e.g.,][]{Dumusque2014}. Some activity indicators, such as FWHM and \sshk, depend only on the projected area of the active regions on the stellar surface, {similar to photometric signals. We refer to these activity indicators as photometric-like \citep[for more details see][]{Isaacson2010,Thompson2017}}. While other quantities such as RV and BIS span {are also sensitive to the change of location of the active regions from the red- to the blue-shifted stellar hemisphere, and vice-versa}
\citep[for a more detailed discussion about this see e.g.,][]{Aigrain2012,Rajpaul2015,pyaneti2}. Therefore, given that the BIS span and RVs depend in a similar way on the stellar activity, it is expected that they present a strong correlation. Therefore, we do not consider correlation between RVs and activity indicators a good proxy to choose the best activity indicator for a multi-GP analysis. 

What we are interested in the multi-GP approach is to find an activity indicator that help us to constrain better the $G(t)$ variable.
We therefore perform a similar analysis to the one presented in Sect.~\ref{sec:multigp} but we use FWHM instead of using the \sshk\ to constrain the shape of the $G(t)$ function. The recovered Keplerian and hyper-parameters are in full agreement with the values reported in Sect.~\ref{sec:dataanalysis}. This is expected given that we foresee that FWHM and \sshk\ constrain the shape of the $G(t)$ function. 

We also performed a 3-dimensional GP regression including the RV, FWHM, and BIS time-series. 
We assume that the BIS span is described by a linear combination of $G(t)$ and $\dot{G}(t)$ \citep[As originally presented by][]{Rajpaul2015}.
Our model setup follows the same guidelines described in Sect.~\ref{sec:multigp}.
As in the previous case, the recovered and Keplerian and hyper-parameters are consistent with the main analysis described in Sect.~\ref{sec:dataanalysis}.
It is worth to mention that for this case we did not see an improvement on the determination of the planetary parameters. This implies that the extra complexity to the model added with the inclusion of the BIS time-series does not help to constrain better the shape of $G(t)$ in this particular dataset.

These results give us confidence that our model to describe the stellar signal in the RV and activity time-series is reliable for this case. We note that we use the \sshk\ time-series in our final model because it is an activity indicator that is independent of the RV extraction method, therefore, it is available for all the instruments.

}

\end{appendix}

\bsp	
\label{lastpage}
\end{document}

%% file: TOI-560-full_params.tex
\newcommand{\smass}[1][$M_{\odot}$]{ $ 0.73 \pm 0.02 $ #1} 
\newcommand{\sradius}[1][$R_{\odot}$]{ $0.65 \pm 0.02 $ #1}

\newcommand{\Tzerob}[1][days]   {$8517.69013 _{ - 0.00059 } ^ { + 0.00056 }$~#1} 
\newcommand{\Pb}[1][d]   {$6.3980420 _{ - 0.0000062 } ^ { + 0.0000067 }$~#1} 
\newcommand{\esinb}[1][ ]   {$-0.03 _{ - 0.18 } ^ { + 0.17 }$~#1} 
\newcommand{\ecosb}[1][ ]   {$-0.22 _{ - 0.15 } ^ { + 0.24 }$~#1} 
\newcommand{\bb}[1][ ]   {$0.566 _{ - 0.068 } ^ { + 0.070 }$~#1} 
 
\newcommand{\rrb}[1][ ]   {$0.03932 _{ - 0.00061 } ^ { + 0.00066 }$~#1} 
\newcommand{\kb}[1][${\rm m\,s^{-1}}$]   {$4.37 _{ - 1.31 } ^ { + 1.46 }$~#1} 
\newcommand{\mpb}[1][$M_{\oplus}$]   {$10.2 _{ - 3.1 } ^ { + 3.4 }$~#1} 
\newcommand{\rpb}[1][$R_{\oplus}$]   {$2.79 \pm 0.10 $~#1} 
 
\newcommand{\eb}[1][ ]   {$0.09 _{ - 0.06 } ^ { + 0.09 }$~#1} 
\newcommand{\wb}[1][deg]   {$-76 _{ - 86 } ^ { + 234 }$~#1} 
\newcommand{\prvb}[1][${\rm km\,s^{-1}}$]   {$6.3 _{ - 6.5 } ^ { + 9.6 }$~#1} 
\newcommand{\ib}[1][deg]   {$88.37 \pm 0.18 $~#1} 
\newcommand{\arb}[1][ ]   {$19.98 _{ - 0.63 } ^ { + 0.61 }$~#1} 
\newcommand{\ab}[1][AU]   {$0.0604 _{ - 0.0026 } ^ { + 0.0027 }$~#1} 
 
\newcommand{\insolationb}[1][${\rm F_{\oplus}}$]   {$43.3 _{ - 4.7 } ^ { + 5.3 }$~#1} 
\newcommand{\tsmb}[1][ ]   {$136 _{ - 35 } ^ { + 60 }$~#1} 
\newcommand{\denstrb}[1][${\rm g\,cm^{-3}}$]   {$3.69 \pm 0.35 $~#1} 
 
\newcommand{\Teqb}[1][K]   {$714 _{ - 20 } ^ { + 21 }$~#1} 
\newcommand{\ttotb}[1][hours]   {$2.143 _{ - 0.027 } ^ { + 0.029 }$~#1}

\newcommand{\denpb}[1][${\rm g\,cm^{-3}}$]   {$2.58 _{ - 0.81 } ^ { + 0.95 }$~#1} 
\newcommand{\grapb}[1][${\rm cm\,s^{-2}}$]   {$1268 _{ - 389 } ^ { + 448 }$~#1} 
\newcommand{\grapparsb}[1][${\rm cm\,s^{-2}}$]   {$1288 _{ - 395 } ^ { + 453 }$~#1} 
 
\newcommand{\Tzeroc}[1][days]   {$9232.1682 _{ - 0.0024 } ^ { + 0.0019 }$~#1} 
\newcommand{\Pc}[1][d]   {$18.87974 _{ - 0.00074 } ^ { + 0.00086 }$~#1} 
\newcommand{\esinc}[1][ ]   {$-0.15 _{ - 0.17 } ^ { + 0.18 }$~#1} 
\newcommand{\ecosc}[1][ ]   {$0.16 _{ - 0.21 } ^ { + 0.17 }$~#1} 
\newcommand{\bc}[1][ ]   {$0.21 _{ - 0.15 } ^ { + 0.23 }$~#1} 
\newcommand{\rrc}[1][ ]   {$0.03368 _{ - 0.00083 } ^ { + 0.00089 }$~#1} 
\newcommand{\kc}[1][${\rm m\,s^{-1}}$]   {$2.89 _{ - 0.51 } ^ { + 0.53 }$~#1} 
\newcommand{\mpc}[1][$M_{\oplus}$]   {$9.7 _{ - 1.7 } ^ { + 1.8 }$~#1} 
\newcommand{\rpc}[1][$R_{\oplus}$]   {$2.39 _{ - 0.09 } ^ { + 0.10 }$~#1} 
 
\newcommand{\ec}[1][ ]   {$0.08 _{ - 0.06 } ^ { + 0.11 }$~#1} 
\newcommand{\wc}[1][deg]   {$-41.6 _{ - 47.7 } ^ { + 52.8 }$~#1} 
\newcommand{\prvc}[1][${\rm km\,s^{-1}}$]   {$-3.1 _{ - 6.3 } ^ { + 3.7 }$~#1} 
\newcommand{\ic}[1][deg]   {$89.72 _{ - 0.27 } ^ { + 0.20 }$~#1} 
\newcommand{\arc}[1][ ]   {$41.11 _{ - 1.3 } ^ { + 1.25 }$~#1} 
\newcommand{\ac}[1][AU]   {$0.1242 _{ - 0.0054 } ^ { + 0.0055 }$~#1}

\newcommand{\insolationc}[1][${\rm F_{\oplus}}$]   {$10.2 _{ - 1.1 } ^ { + 1.2 }$~#1} 
\newcommand{\tsmc}[1][ ]   {$62 _{ - 11 } ^ { + 15 }$~#1}

\newcommand{\Teqc}[1][K]   {$498 \pm 15 $~#1} 
\newcommand{\ttotc}[1][hours]   {$3.67 _{ - 0.102 } ^ { + 0.09 }$~#1}

\newcommand{\denpc}[1][${\rm g\,cm^{-3}}$]   {$3.88 _{ - 0.80 } ^ { + 0.91 }$~#1} 
\newcommand{\grapc}[1][${\rm cm\,s^{-2}}$]   {$1632 _{ - 310 } ^ { + 338 }$~#1} 
\newcommand{\grapparsc}[1][${\rm cm\,s^{-2}}$]   {$1658 _{ - 317 } ^ { + 344 }$~#1} 
 
\newcommand{\qonetess}[1][]   {$0.35 _{ - 0.16 } ^ { + 0.23 }$~#1} 
\newcommand{\qtwotess}[1][]   {$0.16 _{ - 0.12 } ^ { + 0.27 }$~#1}

\newcommand{\qonengts}[1][]   {$0.64 _{ - 0.28 } ^ { + 0.24 }$~#1} 
\newcommand{\qtwongts}[1][]   {$0.37 _{ - 0.23 } ^ { + 0.22 }$~#1}

\newcommand{\qonelcozs}[1][]   {$0.33 _{ - 0.2 } ^ { + 0.32 }$~#1} 
\newcommand{\qtwolcozs}[1][]   {$0.37 _{ - 0.26 } ^ { + 0.33 }$~#1}

\newcommand{\qonelcoB}[1][]   {$0.23 _{ - 0.17 } ^ { + 0.31 }$~#1} 
\newcommand{\qtwolcoB}[1][]   {$0.33 _{ - 0.24 } ^ { + 0.36 }$~#1}

\newcommand{\qonepest}[1][]   {$0.51 _{ - 0.32 } ^ { + 0.34 }$~#1} 
\newcommand{\qtwopest}[1][]   {$0.31 _{ - 0.22 } ^ { + 0.26 }$~#1}

\newcommand{\qonecheops}[1][]   {$0.29 _{ - 0.12 } ^ { + 0.17 }$~#1} 
\newcommand{\qtwocheops}[1][]   {$0.47 _{ - 0.30 } ^ { + 0.31 }$~#1}

\newcommand{\HARPSRV}[1][${\rm km\,s^{-1}}$]   {$20.75208 _{ - 0.00043 } ^ { + 0.00043 }$~#1} 
\newcommand{\HIRESRV}[1][${\rm km\,s^{-1}}$]   {$0.0066 _{ - 0.0023 } ^ { + 0.0021 }$~#1} 
\newcommand{\PFSRV}[1][${\rm km\,s^{-1}}$]   {$-0.00319 _{ - 0.00093 } ^ { + 0.0008 }$~#1} 
\newcommand{\CORALIERV}[1][${\rm km\,s^{-1}}$]   {$20.7371 _{ - 0.0029 } ^ { + 0.0029 }$~#1} 
\newcommand{\HARPSSHK}[1][${\rm km\,s^{-1}}$]   {$0.8793 _{ - 0.0073 } ^ { + 0.0076 }$~#1} 
\newcommand{\HIRESSHK}[1][${\rm km\,s^{-1}}$]   {$0.6369 _{ - 0.0096 } ^ { + 0.0093 }$~#1} 
\newcommand{\PFSSHK}[1][${\rm km\,s^{-1}}$]   {$0.4491 _{ - 0.0098 } ^ { + 0.0099 }$~#1} 
\newcommand{\CORALIESHK}[1][${\rm km\,s^{-1}}$]   {$0.7265 _{ - 0.0099 } ^ { + 0.0097 }$~#1} 
\newcommand{\jHARPSRV}[1][${\rm m\,s^{-1}}$]   {$2.05 _{ - 0.51 } ^ { + 0.59 }$~#1} 
\newcommand{\jHIRESRV}[1][${\rm m\,s^{-1}}$]   {$5.84 _{ - 1.59 } ^ { + 3.04 }$~#1} 
\newcommand{\jPFSRV}[1][${\rm m\,s^{-1}}$]   {$1.87 _{ - 0.44 } ^ { + 0.64 }$~#1} 
\newcommand{\jCORALIERV}[1][${\rm m\,s^{-1}}$]   {$2.62 _{ - 2.1 } ^ { + 4.65 }$~#1} 
\newcommand{\jHARPSSHK}[1][${\rm m\,s^{-1}}$]   {$17.15 _{ - 1.66 } ^ { + 1.91 }$~#1} 
 
\newcommand{\jPFSSHK}[1][${\rm m\,s^{-1}}$]   {$28.31 _{ - 4.15 } ^ { + 5.26 }$~#1} 
\newcommand{\jCORALIESHK}[1][${\rm m\,s^{-1}}$]   {$20.95 _{ - 10.84 } ^ { + 9.71 }$~#1} 
\newcommand{\jtrtess}[1][]   {$0.0006782 _{ - 7.8e-06 } ^ { + 7.9e-06 }$~#1} 
\newcommand{\jtrngts}[1][]   {$0.0037 _{ - 7e-05 } ^ { + 6.7e-05 }$~#1} 
\newcommand{\jtrlcozs}[1][]   {$0.001665 _{ - 4e-05 } ^ { + 4.4e-05 }$~#1} 
\newcommand{\jtrlcoB}[1][]   {$0.00126 _{ - 6.9e-05 } ^ { + 7.2e-05 }$~#1} 
\newcommand{\jtrpest}[1][]   {$0.00338 _{ - 0.00014 } ^ { + 0.00014 }$~#1} 
\newcommand{\jtrcheops}[1][]   {$0.00045 _{ - 2.6e-05 } ^ { + 2.8e-05 }$~#1} 
\newcommand{\jAzero}[1][]   {$0.53 _{ - 0.33 } ^ { + 0.46 }$~#1} 
\newcommand{\jAone}[1][]   {$17.3 _{ - 2.6 } ^ { + 3.1 }$~#1} 
\newcommand{\jAtwo}[1][]   {$0.022 _{ - 0.0034 } ^ { + 0.0041 }$~#1} 
 
\newcommand{\jlambdae}[1][]   {$27.28 _{ - 4.35 } ^ { + 7.26 }$~#1} 
\newcommand{\jlambdap}[1][]   {$0.467 _{ - 0.038 } ^ { + 0.042 }$~#1} 
\newcommand{\jPGP}[1][]   {$12.08 \pm 0.11 $~#1} 

%% file: harpsshort.tex
\begin{table*}
\begin{center}
\caption{HARPS spectroscopic measurements. The full version of this table is available in machine-readable format as part of the supplementary material. \label{tab:harps}}
\begin{tabular}{ccccccc}
\hline\hline
Time & RV & $\sigma_{\rm RV}$ & FWHM & BIS & \sshk\ & $\sigma_{\rm S_{HK}}$ \\
${\rm BJD_{TDB}}$ - 2\,450\,000 & \kms & \kms & \kms & \kms &  &  \\
\hline
8597.587613 & 20.7584 & 0.0021 & 7.0315 & 0.0482 & 0.9146 & 0.0156  \\ 
8601.599930 & 20.7474 & 0.0012 & 6.9998 & 0.0537 & 0.8673 & 0.0089  \\ 
8611.537737 & 20.7505 & 0.0014 & 7.0327 & 0.0727 & 0.8861 & 0.0094  \\ 
8613.520022 & 20.7464 & 0.0013 & 7.0034 & 0.0513 & 0.8992 & 0.0086  \\ 
$\cdots$ \\
\hline
\end{tabular}
\end{center}
\end{table*}

%% file: hiresshort.tex
\begin{table*}
\begin{center}
\caption{HIRES spectroscopic measurements. The full version of this table is available in machine-readable format as part of the supplementary material. \label{tab:hires}} 
\begin{tabular}{cccccc}
\hline\hline
Time & RV & $\sigma_{\rm RV}$ &  \sshk\ & $\sigma_{\rm S_{HK}}$ \\
${\rm BJD_{TDB}}$ - 2\,450\,000 & \kms & \kms &  &  \\
\hline
8777.117606 & 0.0128 & 0.0009 & 0.6552 & 0.0010   \\ 
8788.120576 & -0.0043 & 0.0009 & 0.6224 & 0.0010   \\ 
8795.078952 & 0.0097 & 0.0010 & 0.6585 & 0.0010   \\ 
$\cdots$ \\
\hline
\end{tabular}
\end{center}
\end{table*}

%% file: coralieshort.tex
\begin{table*}
\begin{center}
\caption{CORALIE spectroscopic measurements. The full version of this table is available in machine-readable format as part of the supplementary material. \label{tab:coralie}} 
\begin{tabular}{cccccccccc}
\hline\hline
Time & RV & $\sigma_{\rm RV}$ & FWHM & $\sigma_{\rm FWHM}$ & BIS & $\sigma_{\rm BIS}$ &   \sshk\ & $\sigma_{\rm S_{HK}}$ \\
${\rm BJD_{TDB}}$ - 2\,450\,000 & \kms & \kms & \kms & \kms & \kms & \kms & &  \\
\hline
7655.887186 & 20.7298 & 0.0052 & 8.5236 & 0.0121 & 0.0179 & 0.0074 & 0.6956  & 0.0117   \\ 
7670.851779 & 20.7324 & 0.0045 & 8.5505 & 0.0121 & 0.0230 & 0.0063 & 0.7036  & 0.0102   \\ 
7688.843085 & 20.7589 & 0.0047 & 8.4764 & 0.0120 & -0.0026 & 0.0066 & 0.7841  & 0.0098   \\ $\cdots$ \\
\hline
\end{tabular}
\end{center}
\end{table*}